\documentclass[twocolumn,english,nobibnotes,nofootinbib]{revtex4-1}
\usepackage[T1]{fontenc}
\usepackage[a4paper]{geometry}
\usepackage{float}
\usepackage{textcomp}
\usepackage{amsmath}
\usepackage{amssymb}
\usepackage{graphicx}
\usepackage{esint}
\usepackage{natbib}
\usepackage{color}

\makeatletter
\@ifundefined{textcolor}{}
{%
 \definecolor{BLACK}{gray}{0}
 \definecolor{WHITE}{gray}{1}
 \definecolor{RED}{rgb}{1,0,0}
 \definecolor{GREEN}{rgb}{0,1,0}
 \definecolor{BLUE}{rgb}{0,0,1}
 \definecolor{CYAN}{cmyk}{1,0,0,0}
 \definecolor{MAGENTA}{cmyk}{0,1,0,0}
 \definecolor{YELLOW}{cmyk}{0,0,1,0}
}

\begin{document}

\title{Lorentz force effects for graphene Aharonov-Bohm interferometers}
\author{A. Mre\'{n}ca-Kolasi\'{n}ska}
\affiliation{AGH University of Science and Technology, Faculty of Physics and
Applied Computer Science,\\
 al. Mickiewicza 30, 30-059 Krak\'ow, Poland}

\author{B. Szafran}
\affiliation{AGH University of Science and Technology, Faculty of Physics and
Applied Computer Science,\\
 al. Mickiewicza 30, 30-059 Krak\'ow, Poland}
\begin{abstract}
We investigate magnetic deflection of currents that flow across the Aharonov-Bohm interferometers defined in graphene.
We consider devices induced by closed $n$-$p$ junctions in  nanoribbons as well as etched  quantum rings. 
The  deflection effects on conductance are strictly correlated with the properties of the ring-localized quasibound states.
The energy of these states, their lifetime and the periodicity of the conductance oscillations  are determined by orientation 
of the current circulating within the interferometer. Formation of high harmonics of conductance at high magnetic field and the role of intervalley scattering are discussed.
\end{abstract}
\maketitle

\section{Introduction}

Mesoscopic and nanosize looped conducting channels -- known as quantum rings (QRs) \cite{Aronov87, Buttiker83, Webb85,Timp87,kvon} -- are the simplest electron interferometers that can be defined in solid.
In coherent transport conditions the QR conductance is determined by superposition of wave functions passing through the arms of the ring. The vector potential introduces relative phase shifts to the wave functions \cite{AB}
which result in the Aharonov-Bohm (AB) conduction oscillations with the period of the magnetic flux quantum $\phi_0=e/h$ threading the ring. 
The shifts appear also when the magnetic field ($B$) is present only within the inner core of the ring that is impenetrable for the electron wave function. 
Nevertheless, for nanosize devices the conductance measurements are usually performed in homogeneous magnetic field which is therefore present in the scattering region. The effect of the magnetic field is a deflection of the average electron trajectory \cite{Sza0,Sza1,MF0,MF1,MF2,MF3,MF4} for currents injected to the device 
and formation of the edge currents in the quantum Hall conditions \cite{DAbanin}.

The AB conductance oscillations were studied for QRs etched in graphene 
by both experiment \cite{eg1,eg2,eg3,eg4,eg5,eg6} and theory \cite{tg1,tg2,tg3,tg4,tg5,tg6,tg8,tg9,tg10,tg11,tg13}.
The purpose of the present paper is to describe an interplay of the magnetic deflection and the conductance oscillations.
We consider the conductance calculated by the Landauer approach for atomistic tight binding Hamiltonian and analyze
the resonant states localized in the quantum ring with the stabilization method \cite{mandelsh}.
 The localized states interfere with the incident  Fermi level wave functions  which leaves traces on the conductance oscillations. 
We indicate that it is possible to distinguish two series
of AB oscillations which correspond to localized states with clockwise and anticlockwise persistent current circulation,
that forms magnetic dipole moment parallel or antiparallel to the external magnetic field, respectively.
The series differ in period and width and we explain that the magnetic deflection is responsible for both effects.  
We find that the deflection produces high harmonics of conductance 
at high $B$. 
In the literature the presence of the high harmonics \cite{kvon,eg4} is considered a signature
of phase coherence length being much larger than  the circumference of the ring, with the period of $\phi_0$/n corresponding
to electrons encircling the ring $n$ times \cite{12,13}. For strongly disodered conductors the Al'tshuler-Aronov-Spivak \cite{spivak} periodicity of $\phi_0/2$ 
dominates over the Aharonov-Bohm $\phi_0$ period.
% zdanie o nowych publikacjach
An appearance of the high harmonics 
for graphene quantum rings at high magnetic fields  was recently observed \cite{eg4} and attributed to reduction of scattering involving electron spin flips at high $B$. 
Here we demonstrate that also in the absence of any dephasing effects the high harmonics are activated by high field. 
Note, that the Aharonov-Bohm oscillations were recently studied for systems of multiple quantum dots connected in parallel, in the context of the conductance  harmonics in the nonlinear
and electron-electron interaction effects \cite{Anjou,Kubo,Bedkihal}.

% uzupelnienie na temat temperatury oraz biasu

For $n$-$p$ junctions that are defined in graphene by external voltages the electron trajectories are deflected in opposite directions at both sides of the junction leading to  snake-orbits \cite{snakes,snakes2,snakes3,snakes4,snakes5,snakes6} and the resulting current confinement along the junction \cite{Ghosh,Cresti,Rickhaus,Yliu}.
For a circular $n$-$p$ junction \cite{snakes6} induced in a graphene ribbon by an external potential -- of a scanning probe in particular -- the AB oscillations appear
due to the coupling of the edge and junction currents \cite{amrenca}.
The series of localized states in interferometers both etched and potential-induced are discussed.
The comparison of the two types of interferometers reveals a role of the intervalley scattering for appearance of the conductance oscillations.

\section{theory}

\begin{figure}[tb!]
 \includegraphics[width=\columnwidth]{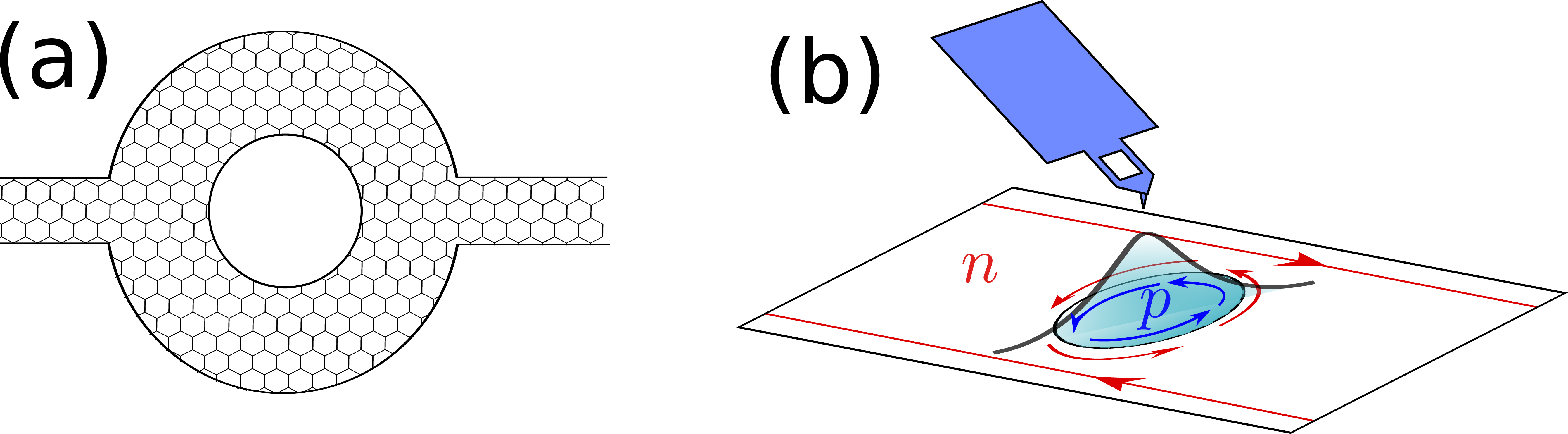}
  \caption{ AB interferometers: a ring etched out of graphene (a) and a circular $n-p$ junction induced within a graphene nanoribbon (b).
  } \label{etch_ind}
\end{figure}

We use the tight-binding Hamiltonian for $\pi$ electrons:
\begin{equation}
   H=\sum_{\langle i,j\rangle }\left(t_{ij} c_i^\dagger c_j+h.c.\right)+\sum_i V({\bf r}_i) c_i^\dagger c_i, 
\label{eq:dh}
\end{equation}
where the first summation runs over the nearest neighbors, 
and $V({\bf r}_i)$ is the external potential at position $\mathbf{r}_i$ of the $i$-th ion.
We consider two types of devices: the etched and the induced ones. The etched rings are  connected to input and output leads by two narrow graphene nanoribbons [see Fig.~\ref{etch_ind} (a)].
 The induced device  consists of a wider graphene nanoribbon with a scanning gate microscope tip floating above [Fig.~\ref{etch_ind} (b)].
For the etched rings the potential  $V({\bf r}_i)$ is taken zero everywhere. For the induced interferometers $V$ describes the effective potential of the tip. 
Due to screening of the Coulomb potential of the tip by the two-dimensional electron gas, we assume effective potential given by a Lorentzian function, according to the Schr\"odinger-Poisson modeling \cite{kolasinskiDFT2013} 
\begin{equation}
V({\bf r})=\frac{V_t}{1+\left(|{\bf r}-{\bf r }_{t}|/d \right)^n}, \label{lf}
\end{equation}
where $n=2$, ${\bf r}_t=(x_t,y_t,0)$ stands for the tip position, $d$ for the width of the effective tip potential, and $V_t$ is the maximal value of the tip potential.

The hopping elements of the first sum in Eq.~(\ref{eq:dh}) include the Peierl's phase,  $t_{ij} = t \exp( \frac{2\pi i}{\phi_0} ) \int_{\mathbf{r}_i}^{\mathbf{r}_j} \mathbf{A}\cdot\mathbf{dl} $, where $t$ is the hopping parameter. The magnetic field is applied perpendicular to the plane of confinement $\mathbf{B}=(0,0,B_0)$, and we use the Landau gauge,  $\mathbf{A}=(-yB_0,0,0)$.
We consider the energy range near the Dirac point. The numerical complexity of the problem can be reduced by the scaling approach of Ref.~\cite{Rickhausprl} that we apply here. The ribbons modeled here are scaled up with the condition $a=a_0 s_f$ and $t=t^0/s_f$, where  $t^0=-2.7$ eV is the unscaled hopping parameter,   $a_0=2.46$ {\AA } is the graphene lattice constant. We apply a scaling factor of $s_f=4$.
The rescaled magnetic field is $B_0=B s_f^2$, with $B$ being the actual magnetic field applied to the modeled sample.

In order to evaluate the transmission probability, we use the wave function matching method (WFM), as described in Ref.~\cite{Kolacha}. The transmission probability from the input lead to mode $m$ in the output lead is
\begin{equation}
T^m = \sum_{ n } |t^{mn}|^2,
\label{eq:transprob}
\end{equation}
with $t^{mn}$ being the probability amplitude for the transmission from the mode $n$ in the input lead to mode $m$ in the output lead. The linear conductance \cite{datta} is evaluated as  $G={G_0}T_{tot}$, with $T_{tot}=\sum_{m} T^{m}$ and $G_0={2e^2}/{h}$.

The current flow between the atoms $m$ and $n$, as derived from the Schr\"odinger equation \cite{Wakabayashi}, is
\begin{equation}
 J_{mn} =  \frac{i}{\hbar} \left[ t_{mn} \Psi^*_m \Psi_n - t_{nm} \Psi^*_n \Psi_m \right],
\label{current}
\end{equation}
where $ \Psi_n $ is the wave function at the $n$th atom.
The probability current flux can be evaluated as:
\begin{equation}
 \phi = \sum\limits_m \sum\limits_{n_m}  J_{mn_m} ,
\label{currentFlux}
\end{equation}
where the first sum runs over the atoms along a cross-section of the ribbon, and the second sum runs over their neighbors $n_m$ localized to the right.

The results provided below are analyzed with respect to the density of the localized resonant states. 
The density is evaluated with the stabilization method \cite{mandelsh} [see the Appendix].

\section{results}

\begin{figure}[tb!]
  \includegraphics[width=\columnwidth]{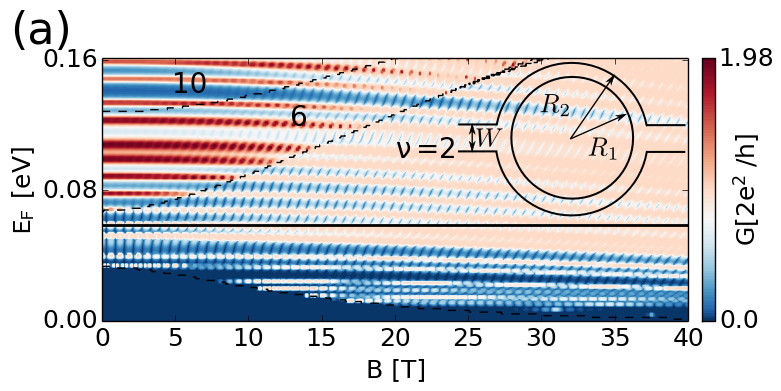}
  \includegraphics[width=\columnwidth]{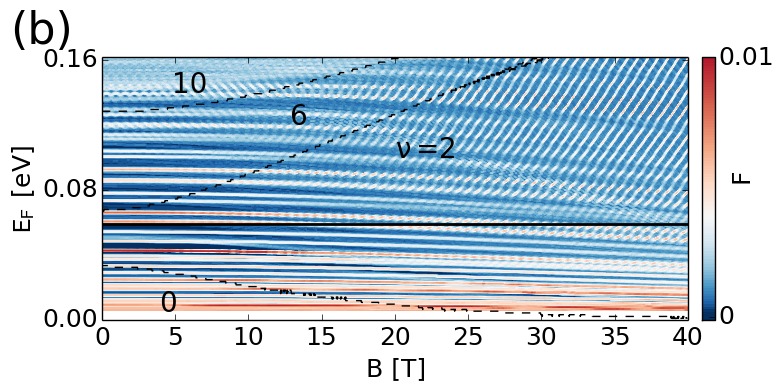}
  \caption{(a) Conductance of a narrow quantum ring ($R_1=41.05$ nm, $R_2=48.95$ nm) 
connected to a semiconducting armchair ribbon of width $W=17.23$ nm. (b) Counter of the
localized states as determined by the stabilization method. Dashed lines separate the regions of varied filling factor $\nu$ in the ribbon. The solid vertical lines indicate the Fermi energies  studied in detail. } \label{nasa}
\end{figure}

\begin{figure}[tb!]
 \includegraphics[width=\columnwidth]{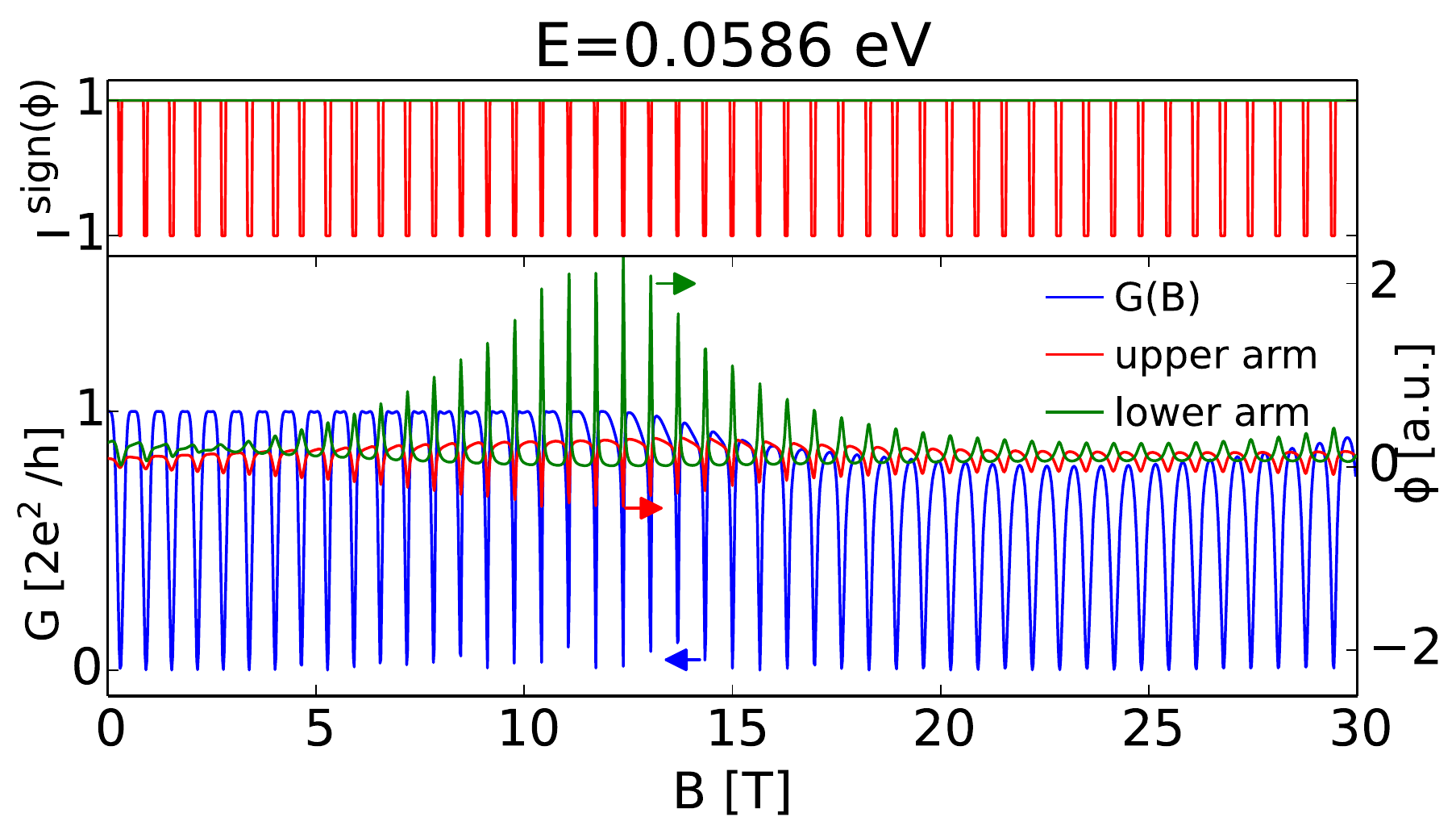}
  \caption{ Conductance (blue line) -- cross section of Fig.~\ref{nasa}(a) for $E=0.0586$ eV, the current flux through the upper and lower arms of the ring (see Fig.~\ref{schem}) and the sign of the flux (upper panel). 
  } \label{strum}
\end{figure}

\begin{figure}[tb!]
  \includegraphics[width=0.5\columnwidth]{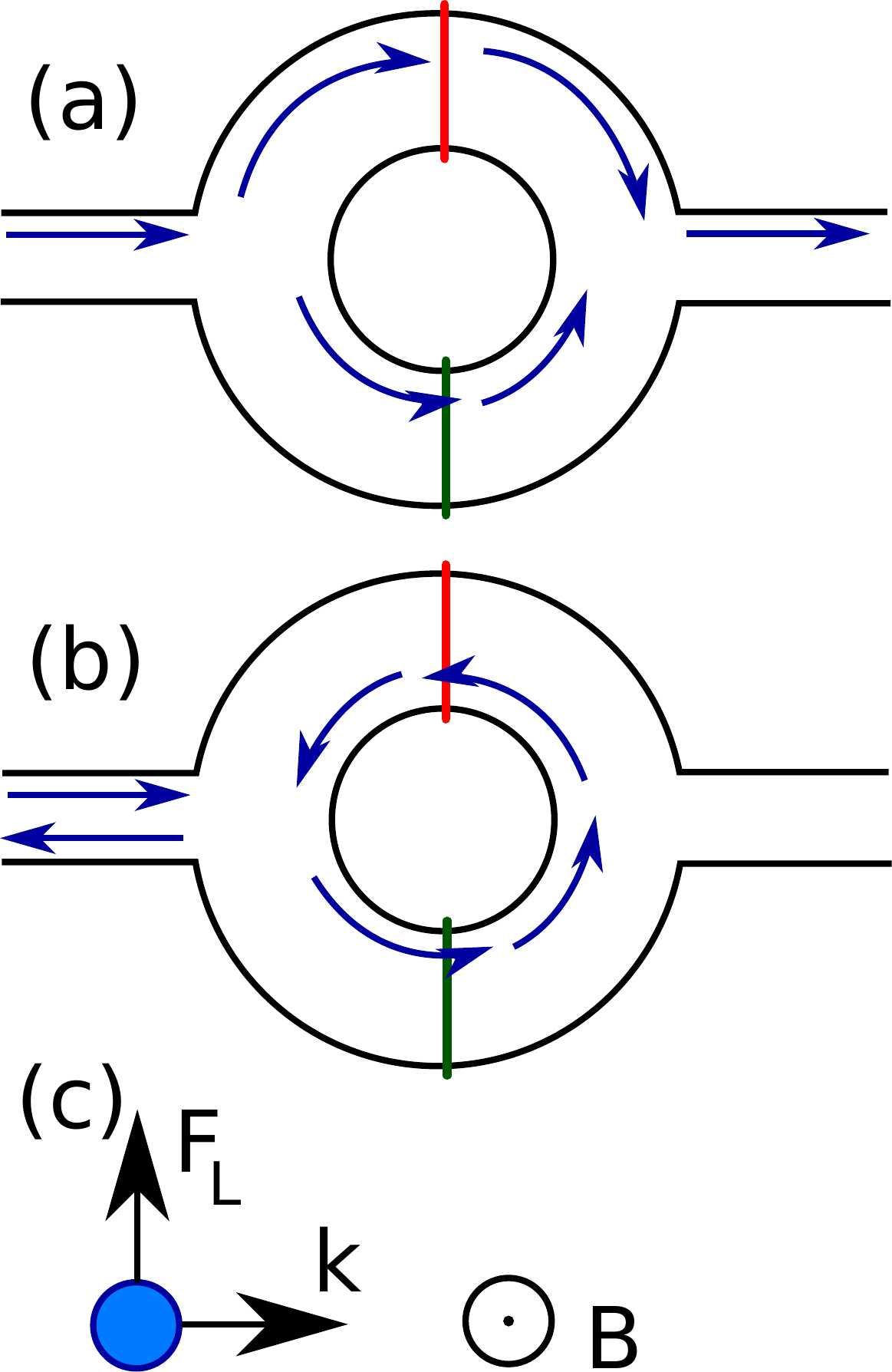}
  \caption{Schematics of the electron flow at the maxima (a) and minima (b) of $G$ in Fig.~\ref{strum} and
(c) the orientation of the Lorentz force for the electron moving right in the out-of-plane magnetic field.} \label{schem}
\end{figure}

\begin{figure}[tb!]
  \includegraphics[width= \columnwidth]{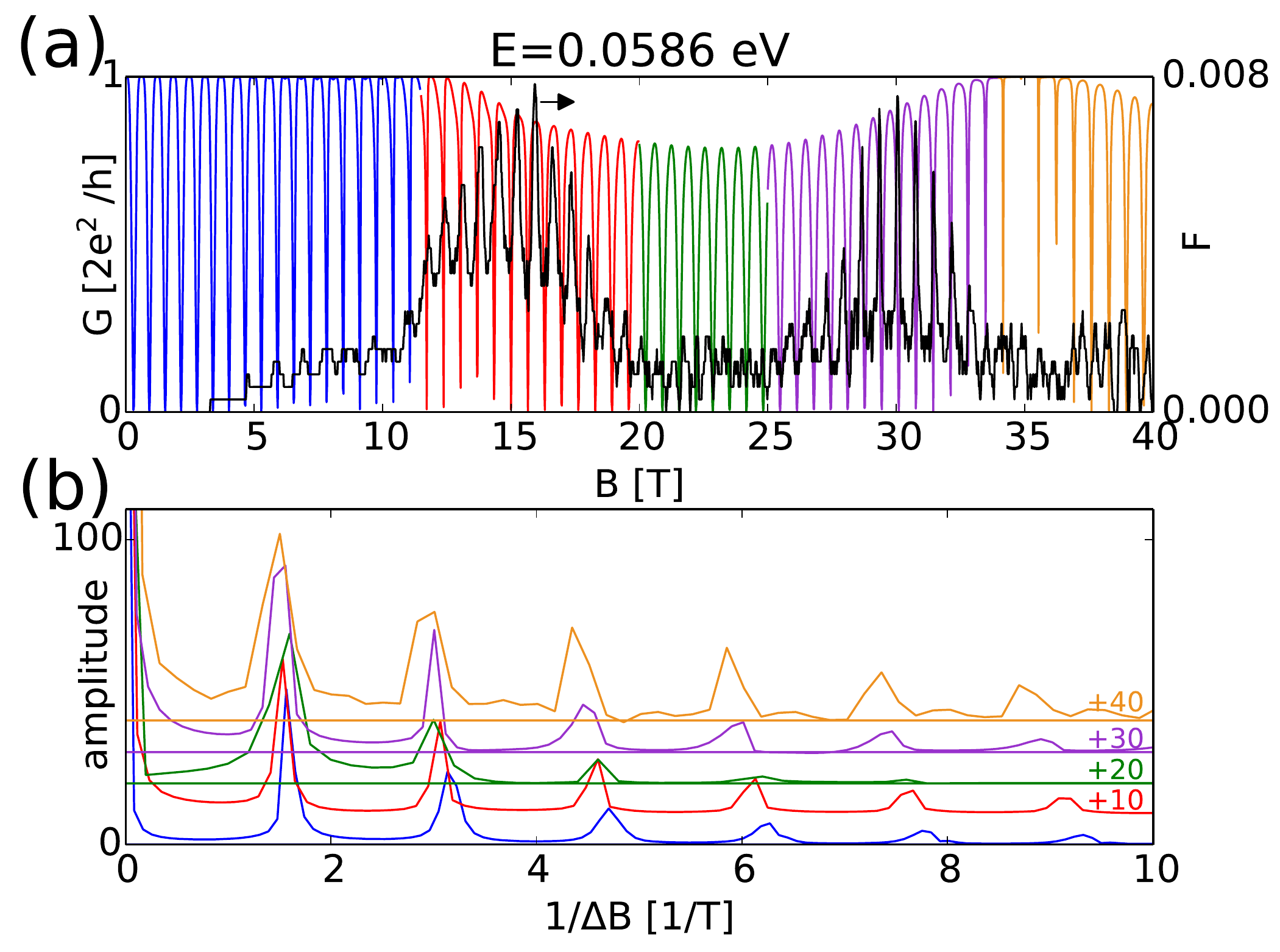}
  \caption{(a) Conductance (line of varied colors) -- same as in Fig.~\ref{strum}, and the resonance counter for $E=0.0586$ eV. (b) Fourier transform calculated for intervals of the magnetic field which marked in the corresponding colors. Each plot in (b) is normalized so that the first peak has the same amplitude as in the blue curve.}\label{fft1}
\end{figure}

\begin{figure}[tb!]
  \includegraphics[width= \columnwidth]{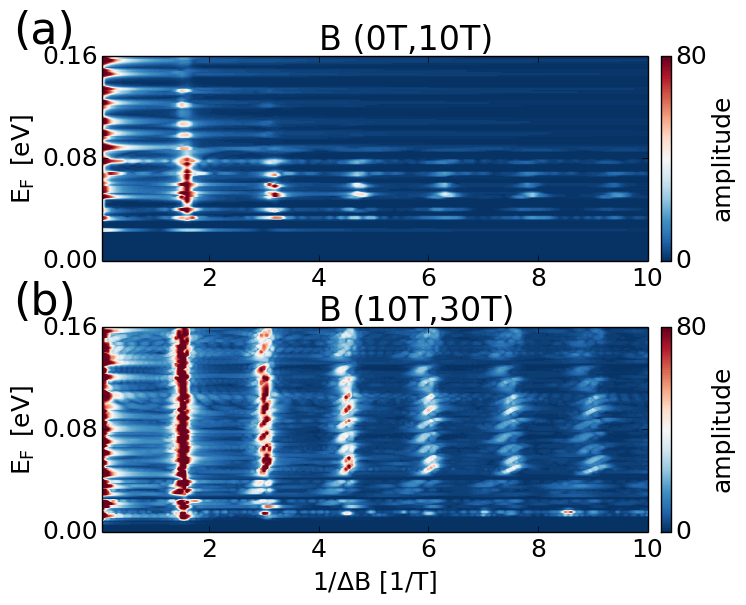}
  \caption{ Fourier transform of Fig.~\ref{nasa}(a) for $B\in(0,10$T) (a), and for $B\in(10$T$,30$T) (b).} \label{fftu}
\end{figure}

\subsection{Narrow etched ring}
Let us begin the discussion by a narrow ring (the internal radius $R_1=41.05$ nm, and the external one $R_2=48.95$ nm -- see the inset to Fig.~\ref{nasa}(a)) for which the simplest periodic behavior in external magnetic field can be expected. 
The leads are semiconducting armchair nanoribbons of width $W=17.23$ nm. 
Both the conductance [Fig.~\ref{nasa}(a)] and the resonance counter [Fig.~\ref{nasa}(b)] indicate
a number of nearly horizontal lines that change only weakly with $B$. These lines correspond to 
states that are localized near the edges of the ring, where short zigzag segments appear (Fig.~\ref{schemstab}(a)).
The strong localization is responsible for weak dependence on $B$. 
For larger magnetic field the stability diagram [Fig.~\ref{nasa}(b)] contains a series of lines growing steep up in the energy as $B$ is increased. The trace of these lines is visible -- although weaker in the conductance plot [Fig.~\ref{nasa}(a)]. 

In order to identify these lines a cross section of Fig.~\ref{nasa}(a) was plotted in blue in Fig.~\ref{strum} for $E=0.0586$ eV [the lowest horizontal line in Fig.~\ref{nasa}(a,b)].
In Fig.~\ref{strum} the conductance is accompanied  by the current fluxes that pass through the upper and lower arms of the ring  \cite{uwaga},
for the current injected to the ring from the left channel [Fig.~\ref{schem}(a)].
The conductance oscillations of Fig.~\ref{strum} are strictly correlated with the periodically changing properties
of the current flow: at the maxima of the transfer probability the current flows
to the right at both the upper and lower arms of the ring [Fig.~\ref{schem}(a)], however the dips of $T$ correspond to a reversal of the current in the upper arm. The current forms then a counterclockwise loop around the ring [Fig.~\ref{schem}(b)]. This current orientation produces the magnetic dipole $\mu$ which is antiparallel to the external magnetic field \cite{szafran}. The interaction of this dipole with the magnetic field ($\Delta E=-\vec{\mu} \cdot \vec{B}$)  leads to the growth of the energies with $B$ that is visible in Fig.~\ref{nasa}(a).
In Fig.~\ref{fft1}(a) the transfer probability -- still for $E=0.0586$ eV is confronted with the localized
resonances counter $F$. For the peaks of the counter the dips of conductance appear, hence the reversal of the current occurs at interference of the incoming electron with the ring localized quasi-bound states.
For the anticlockwise current circulation the Lorentz force keeps the current confined within the ring [Fig.~\ref{schem}(c)] hence the very pronounced %of the 
localized resonances in the stabilization diagram [Fig.~\ref{nasa}(b)].
On the other hand, the localized states that correspond to the opposite current circulation are destabilized by the Lorentz force, so they leave only a trace on the stability diagram
of Fig.~\ref{nasa}(b) -- but still they can be noticed at the high energy region in the upper right corner of the plot. In graphene the cyclotron radius for the Fermi energy $E_F$   \cite{MF2} is $R_c=\frac{E_F}{V_F eB}$, where $V_F$ is a
material constant. For higher $E_F$ the cyclotron radius becomes larger than the size of the ring-ribbon junction, which reduces the magnetic deflection that for this current orientation
tends to eject the resonance states out of the ring and hence delocalize the resonant states.

Figure \ref{fft1}(b) shows the Fourier transform of $G(B)$ calculated in finite ranges of $B$ that are marked with different colors in Fig.~\ref{fft1}(a). The plots of Fourier transform are normalized so that for each plot the first peak has the same amplitude as in the blue curve in Fig. \ref{fft1}(b). We find two distinct features: (i)  the period of the resonances increases with $B$ and (ii) the higher harmonics are enhanced at high magnetic field. Both these features can be explained as consequences of the Lorentz force.
Feature (i) is due to a reduced effective radius of the ring which results from the Lorentz force pushing the wave function to the inner core of the ring [cf. Fig.~\ref{schem}(b)], hence the AB period $\Delta B=\frac{h}{eA} = \frac{2\hbar}{eR^2}$ increases. The feature (ii) results from the stabilization of the anticlockwise loop inside the ring that increases the number of turns of the electron circulation around the ring.

The results discussed in Figs. \ref{nasa}, \ref{strum} and \ref{fft1} were obtained for the lowest subband transport.  For higher filling factors -- when the intersubband scattering is present --  the  AB oscillation becomes pronounced only for $B$ larger than 10T  [Fig.~\ref{nasa}(a)] , i.e. when the resonant lines are formed in Fig.~\ref{nasa}(b). 
The results for high harmonics are summarized in Fig.~\ref{fftu} which shows the Fourier transform of the signal for varied energies within the range of (0,10)T [Fig.~\ref{fftu}(a)] and (10,30)T  [Fig.~\ref{fftu}(b)]. Also at higher energies the magnetic field enhances the higher harmonics. The stabilization of the higher harmonics in the stronger magnetic fields
was recently found in Ref.~\onlinecite{eg4}.

\begin{figure}[tb!]
  \includegraphics[width= \columnwidth]{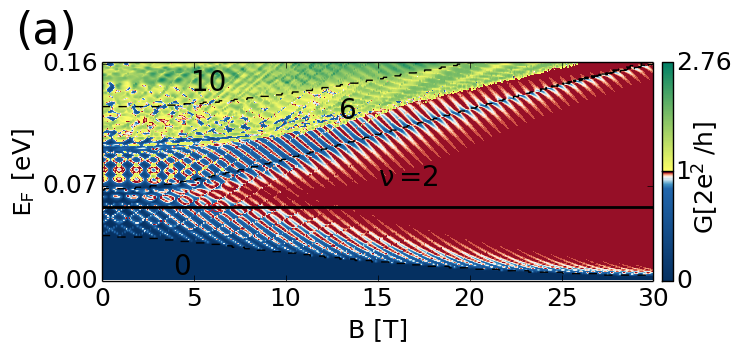}
\includegraphics[width= \columnwidth]{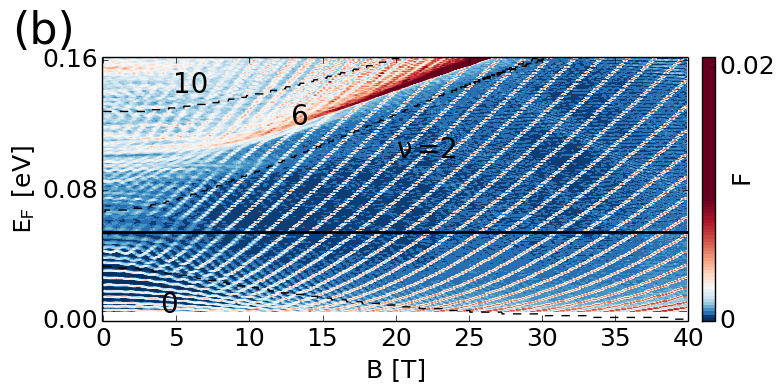}
  \caption{Same as Fig.~\ref{nasa} only for a wider ring to $R_1=23.4$ nm and $R_2=48.95$ nm. } \label{wide}
\end{figure}

\begin{figure}[tb!]
  \includegraphics[width= \columnwidth]{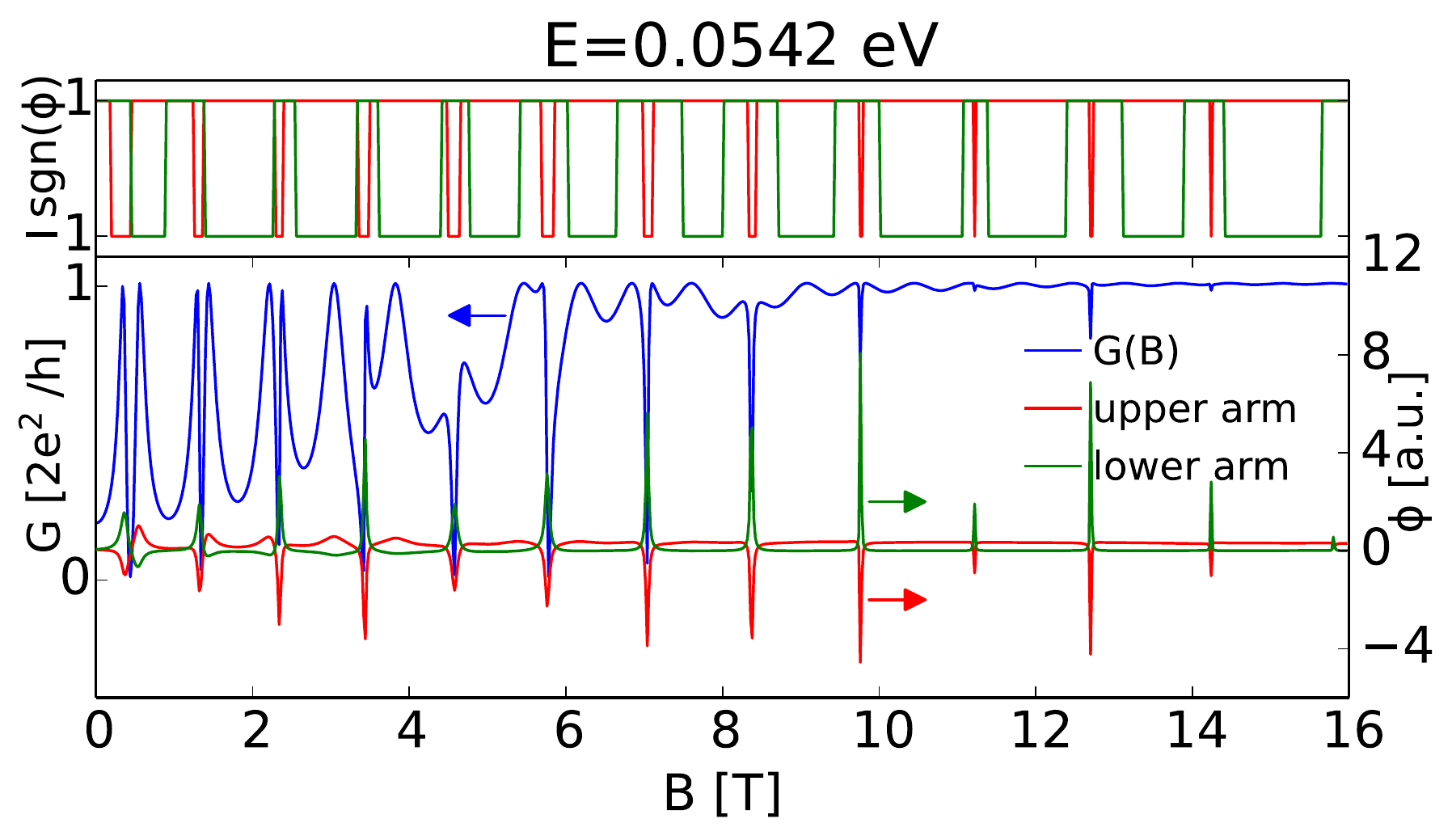}
  \caption{Same as Fig.~\ref{strum} only for a wider ring to $R_1=23.4$ nm and $R_2=48.95$ nm. The conductance plotted in blue is a cross section  of Fig.~\ref{wide} for $E=0.0542$ eV.
} \label{wS}
\end{figure}

\begin{figure}[tb!]
  \includegraphics[width=0.48\columnwidth]{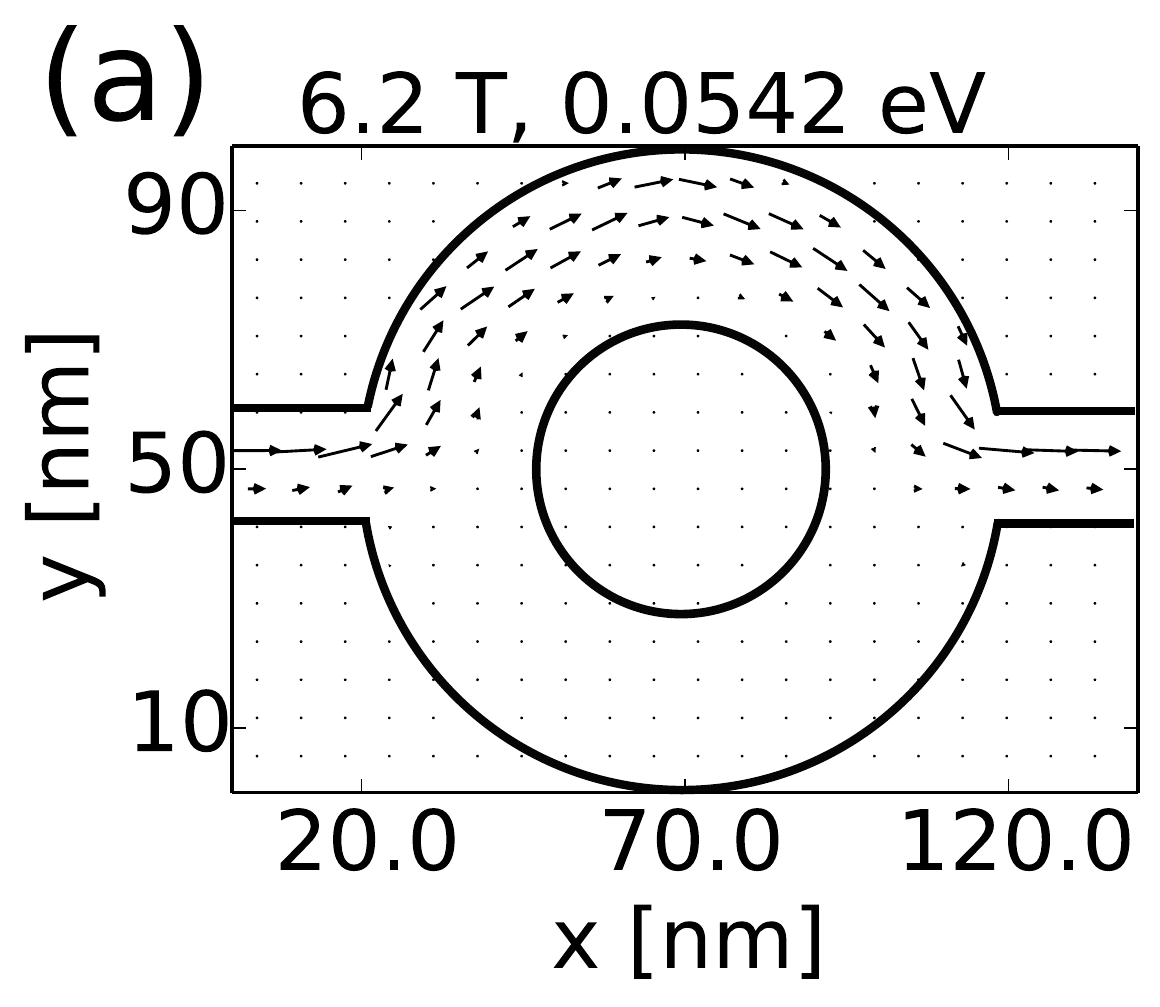}
  \includegraphics[width=0.48\columnwidth]{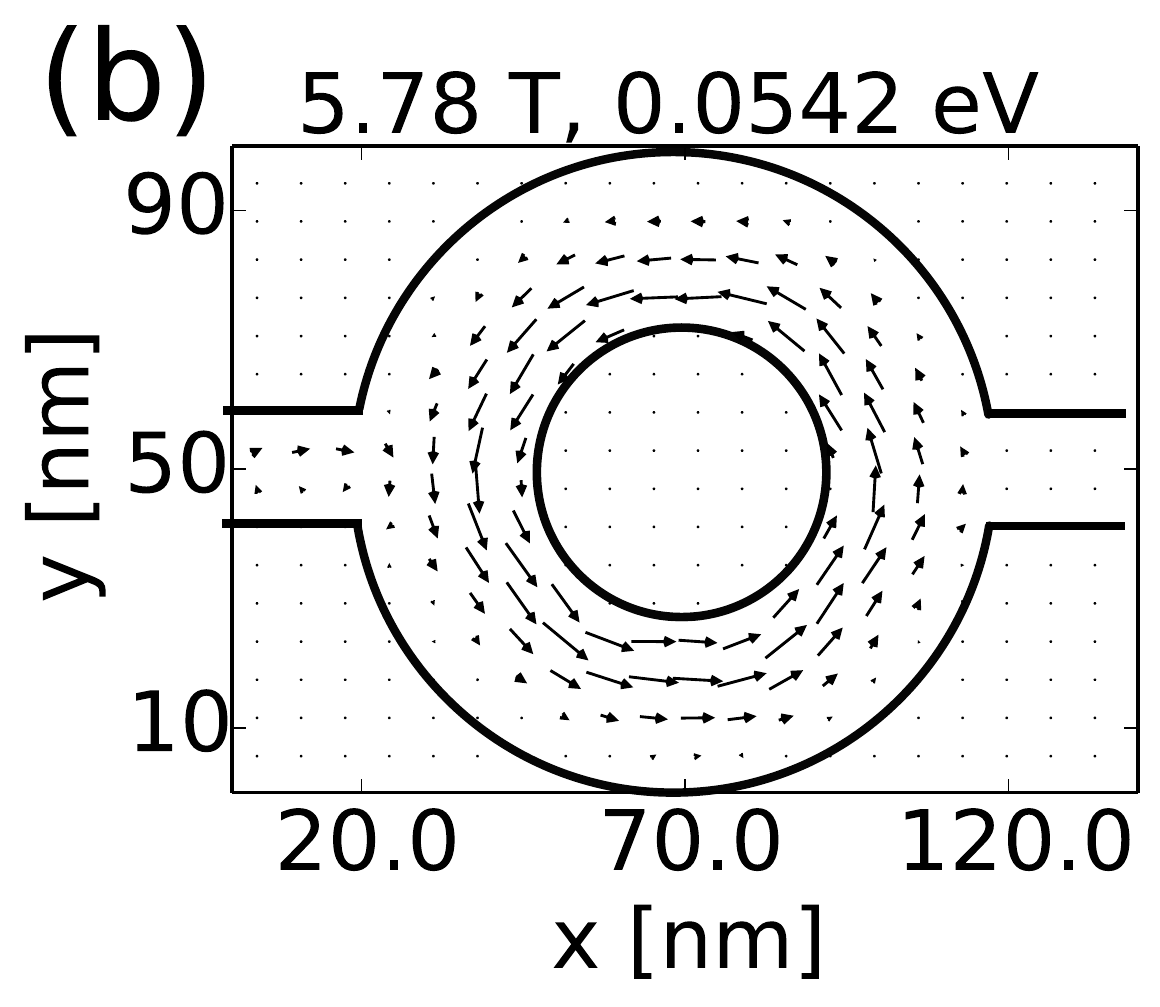}
  \caption{The current distribution for the results of Fig.~\ref{wS} for (a) $B=6.2$ T (a peak of conductance) and (b) for $B=5.78$ T (a dip of conductance.)} \label{cc}
\end{figure}

\subsection{Wide etched ring}
The magnetic deflection effects discussed above were limited by the narrow width of the ring. In this part we change the ring radii to $R_1=23.4$ nm and $R_2=48.95$ nm. 
The results for conductance and the resonance counter are given in Fig.~\ref{wide}. For $B\in(0,10)$T one observes both the lines which grow and decrease in the energy
with $B$ -- due to the resonant states of both current orientations. Above 10T the conductance plot [Fig.~\ref{wide}(a)] resolves only the states which go down in the energy (clockwise currents) with an increase of $B$, and in contrast -- the stability diagram [Fig.~\ref{wide}(b)]  retains only the states that increase in the energy with growing $B$.

The cross section of Fig.~\ref{wide} is given in Fig.~\ref{wS} and shows that at high magnetic field wide peaks and narrow dips of conductance appear.
The wide peaks correspond to a clockwise current circulation [Fig.~\ref{cc}(a)] while the narrow dips appear with an anticlockwise current [Fig.~\ref{cc}(b)].
The shifts of the peaks [Fig.~\ref{wide}(a)] and dips [Fig.~\ref{wide}(b)] in energy agree with the orientation of the produced dipole moment with respect to the external magnetic field. 
At high $B$ -- the dips become too thin to be resolved on the conductance plot [Fig.~\ref{wS}] and the peaks extend to almost any magnetic field.
The width of the resonances and antiresonances is related to the lifetime of the quasibound states. 
 The lifetime of the dip-related resonances becomes very large -- as they become decoupled from the 
states of the channel -- and disappear from the conductance plot of Fig.~\ref{wide}(a). On the other hand, the lifetime of the states with the opposite current circulation becomes very small
which removes them from the stability plot of Fig.~\ref{wide}(b). 

In addition,  in Fig.~\ref{wide}(a,b) we can see that below 10T both the lines that grow and decrease with $B$ appear at same energy with the same period in the external magnetic field.
However, for higher magnetic field the states that decrease with $B$ in Fig.~\ref{wide}(a) appear with a much shorter period than the ones that grow in the energy in Fig.~\ref{wide}(b). 
The change of the periodicity is related to the wave function being shifted to the internal or external edge of the ring depending on the current circulation and the orientation of the Lorentz force.
The opposite shifts for opposite current circulation are well visible in Fig.~\ref{cc}.

\begin{figure}[tb!]
  \includegraphics[width=0.7\columnwidth]{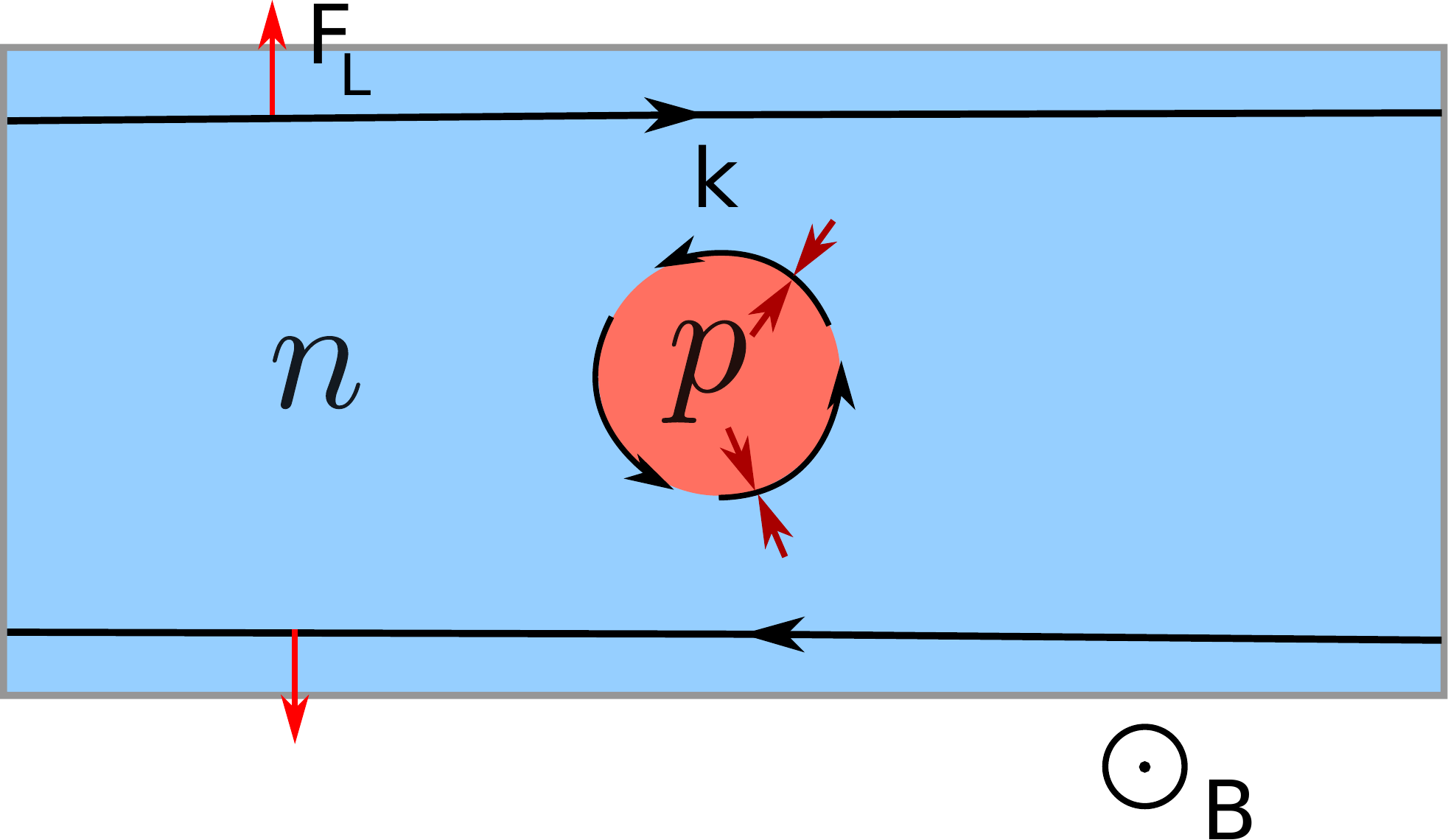}
  \caption{Schematics of the electron flow for the AB interferometer induced within an $n$-type nanoribbon by external potential 
forming a circular $p$-type region in the center of the ribbon. The black arrows indicate the electron currents and the red arrows
the direction of the Lorentz force for out-of-plane magnetic field. In the simulation the electron is incident from the left
near the upper edge. } \label{schnp}
\end{figure}

\begin{figure}[tb!]
  \includegraphics[width= \columnwidth]{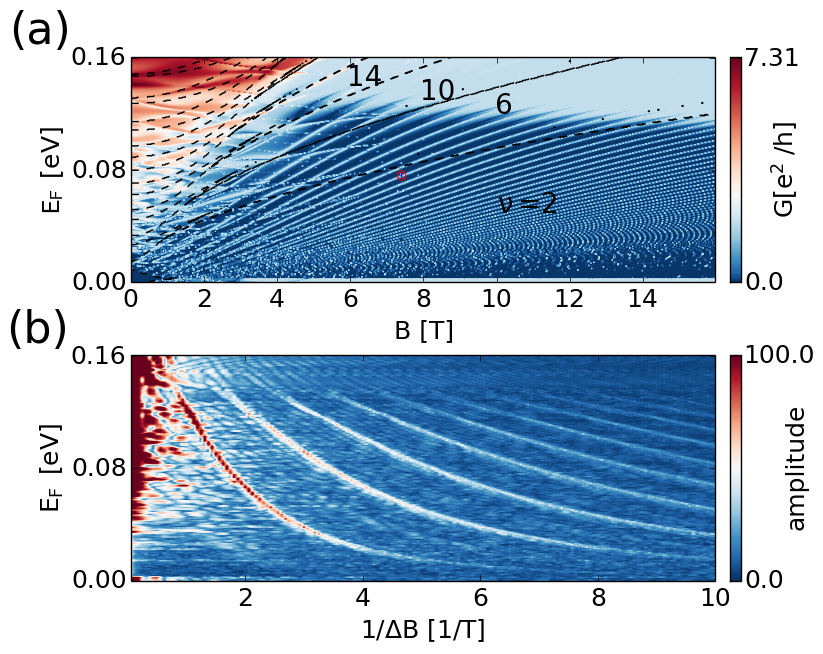}
\includegraphics[width= \columnwidth]{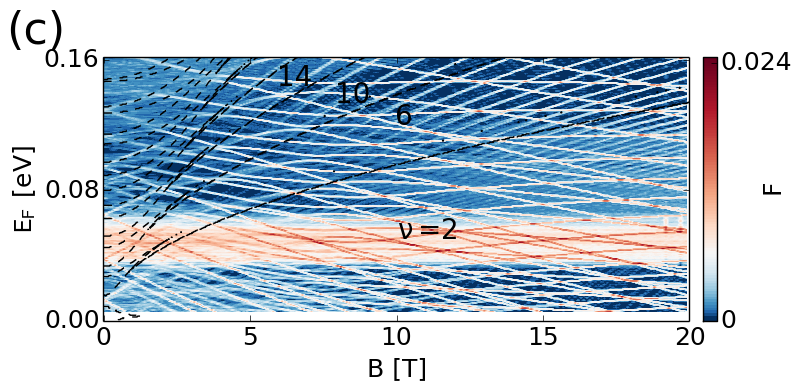}
  \caption{Conductance of the $n$-$p$ inteferometer (a), its Fourier transform (b), and the resonance counter $F$ (c). } \label{tip}
\end{figure}

\begin{figure}[tb!]
  \includegraphics[width=0.7\columnwidth]{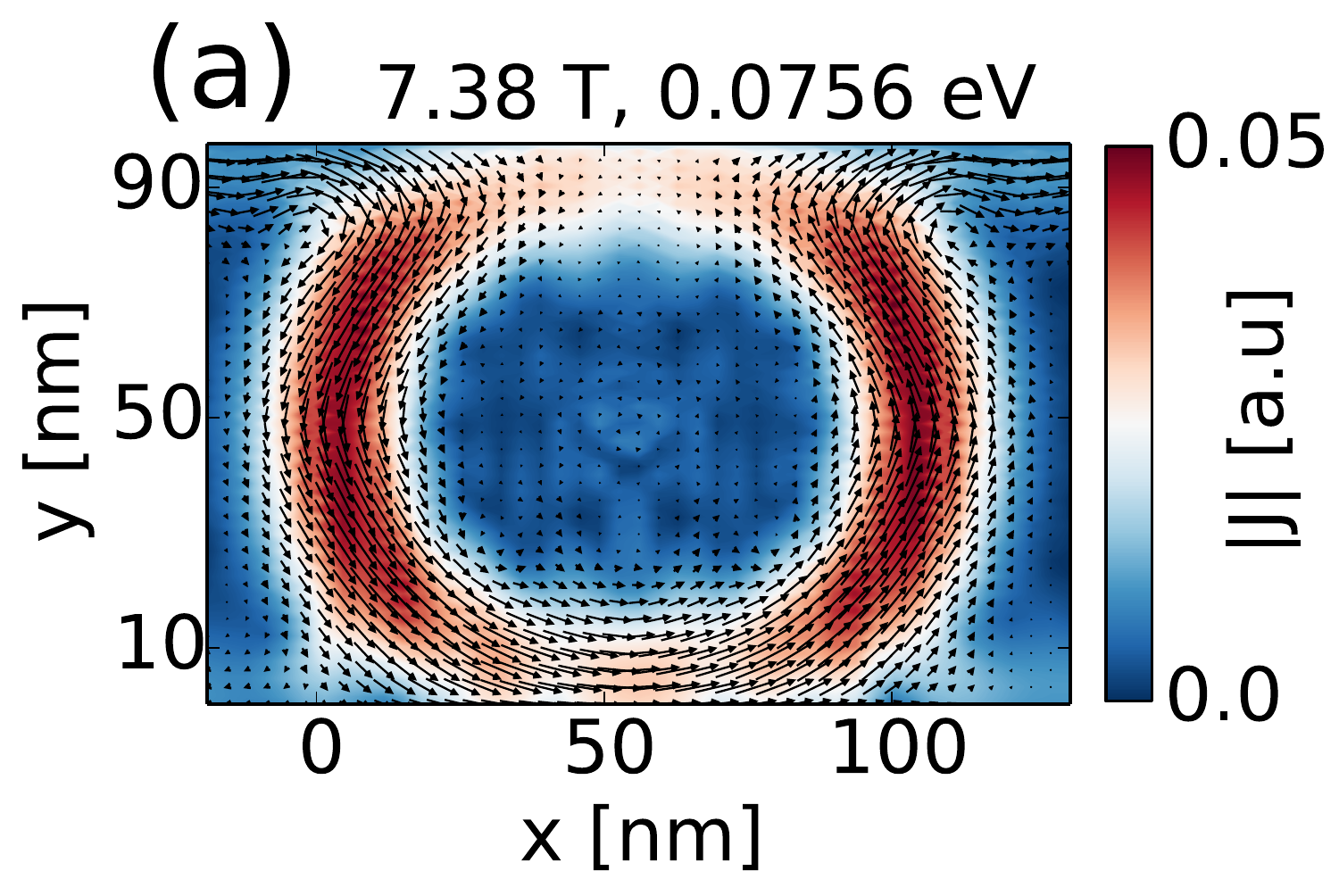}
\includegraphics[width=0.7\columnwidth]{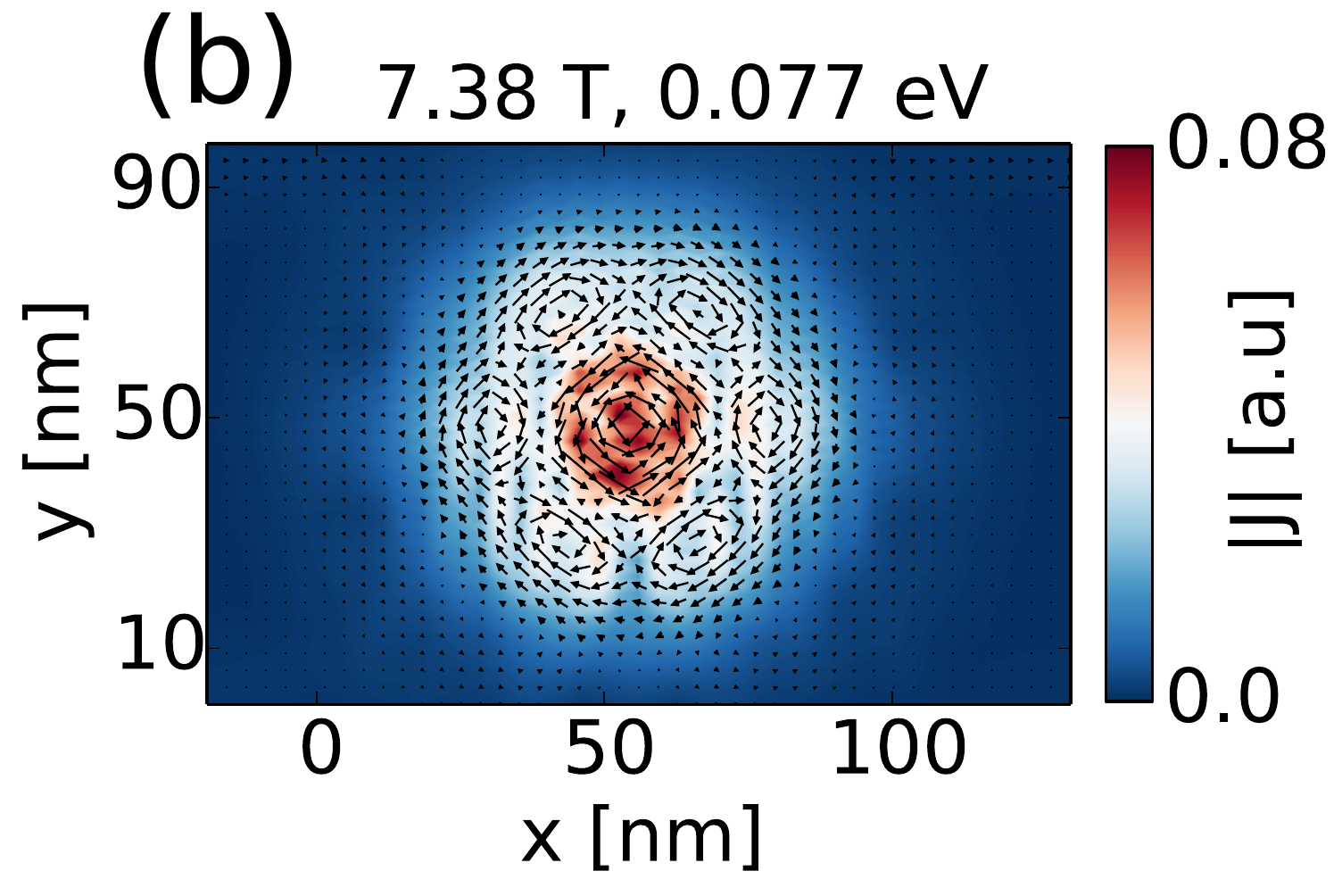}
  \caption{The electron density obtained from the diagonalization of a closed system (as in the stabilization algorithm) and the current distribution
for (a)  one of the peaks of conductance ($E=0.0756$ eV, $B=7.38$T)with the circulation of the current along the $n$-$p$ junction and for one of the resonances formed inside the $p$ region ($E=0.077$ meV, $B=7.38$ T).   } \label{tipr}
\end{figure}

\subsection{Induced quantum rings}

%The AB interferometers taylored of the graphene plane by etching resonances of both current circulation were present. 
Let us now briefly discuss an AB interferometer that is formed in a graphene nanoribbon by external potential
of e.g. a scanning gate microscope placed near the center of the ribbon. We consider $E_F>0$ and $V_F>E_F$
so that the tip forms a $p$-type conductivity region (see Fig.~\ref{schnp}) within the $n$-type nanoribbon \cite{amrenca}. This $n$-$p$ junction 
stabilizes an anticlockwise current \cite{Yliu} with the Lorentz force acting towards the junction on its both sides (red arrows in Fig.~\ref{schnp}).

For numerical calculations we took the armchair nanoribbon of width $W=98.4$ nm, and the parameters of the potential induced by the tip were $ V_t=0.4 $ eV and $ d=24.6 $ nm.
In the conductance dependence on the magnetic field only the resonances that increase with external magnetic field are observed [cf. Fig.~\ref{tip}(a)]. 
The current distribution -- calculated from the transport problem -- and the electron density at the resonance  -- calculated by diagonalization
of the closed version of the system -- is given in Fig.~\ref{tipr}(a), which shows the passage of the edge currents around the junction. 

The Fourier transform of conductance [Fig.~\ref{tip}(b)] indicates the presence of the higher harmonics. In contrast 
to the results for the etched rings, the oscillation period increases with the Fermi energy -- since the diameter of the 
$p$ region is reduced to zero when the Fermi energy is increased (in the limit of $E_F=V$ the $p$ region disappears).
The stability diagram [Fig.~\ref{tip}(c)] contains the resonances that are present in the conductance dependence [see the higher energy part of Fig.~\ref{tip}(c) at high $B$],
however they appear rather in the background of the plot. On the foreground there is a multitude of lines separated by larger distances than the transport resonances.
These new lines  at high $B$ have a very high contrast and (i) either decrease or (ii) increase slowly in the energy  with $B$.
These lines correspond to states which are confined entirely within the $p$ potential -- under the tip, which (i) form an clockwise current loop (for the ''paramagnetic'' states
that decrease in external magnetic field) or (ii) produce a weak magnetic dipole with anticlockwise current that is induced by the external magnetic field (''diamagnetic'' states).
An example of the latter is displayed in Fig.~\ref{tipr}(b). The resonant states
localized under the tip are screened from the edge currents by the anticlockwise loop formed at the $n$-$p$ interface.

 Note, the sharp resonances found in the conductance of Fig. \ref{tip} could be used as a sensor of the magnetic field  inhomogeneity with the resolution of the sensor given by the width of the conductance peak.

\subsection{Zigzag leads}

\begin{figure}[tb!]
  \includegraphics[width= \columnwidth]{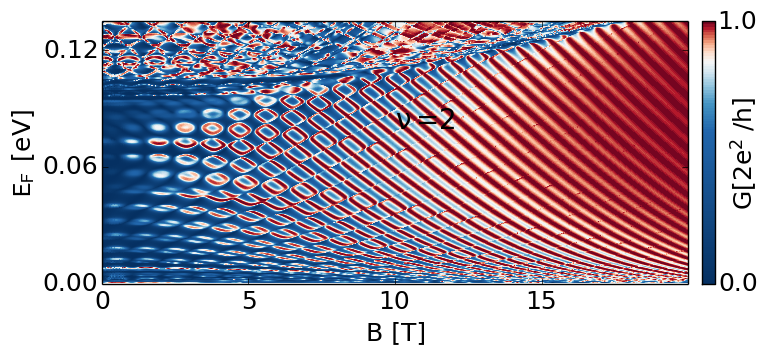}
  \caption{Same as Fig.~\ref{wide}(a) only for the ring connected to a zigzag ribbon. } \label{zz}
\end{figure}

\begin{figure}[tb!]
  \includegraphics[width= \columnwidth]{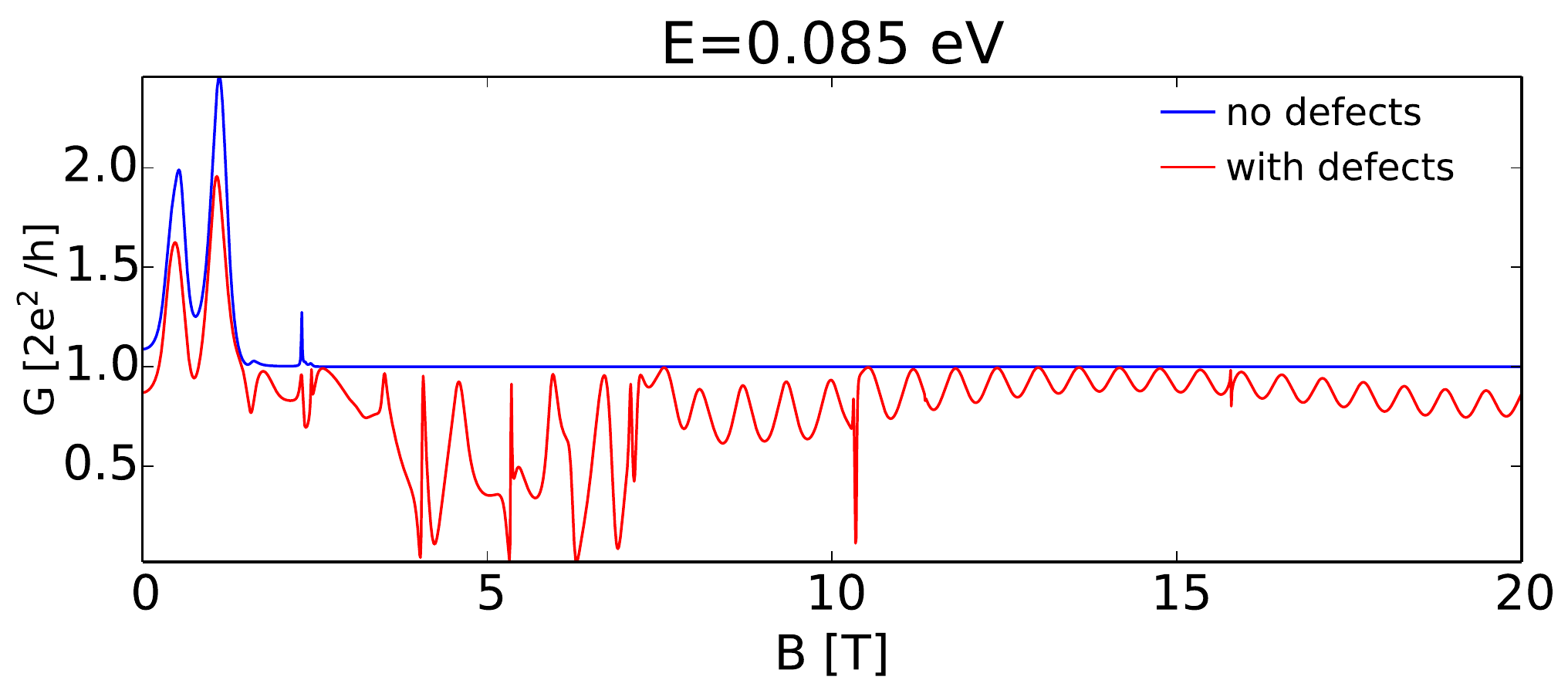}
  \caption{Conductance for the circular $n$-$p$  junction defined within a zigzag ribbon without (blue line) and with (red line) edge defects (see text). }\label{zzt}
\end{figure}

For an etched ring connected to zigzag ribbons acting as leads (conductance given in Fig.~\ref{zz}) with the internal radius $R_1=23.3$ nm, the external one $R_2=49.1$ nm, and leads 17.6 nm wide, 
one obtains very similar results to the one obtained above for the semiconducting armchair ribbon [Fig.~\ref{wide}(a)]: with 
resonances of both current orientations visible at low field and only the clockwise surviving at high field.  A difference with respect to Fig.~\ref{wide}(a) is the non-zero conductance
for low $E$ that is observed only for the zigzag ribbon -- due to their metallic character.
However, an attempt to form the  AB interferometer induced within the zigzag ribbon by an external potential apparently fails --
see the blue line in Fig.~\ref{zzt}. No AB-periodic conductance oscillations are formed at high field.

The reason for the absence of the AB oscillation for the interferometer induced in the zigzag ribbon is the absence of backscattering  at $\nu=2$.
The lack of backscattering \cite{pcc,pccc} in the lowest subband for the zigzag ribbon results from the chiral character of the carriers,
or -- in other words -- the fact that the backscattering would require an intervalley scattering and the latter cannot be induced by a smooth, long-range
external potential forming the $p$ region inside the ribbon \cite{pcc,pccc}. For the etched structure [Fig.~\ref{zz}] the edges of the ring (Fig.~\ref{schemstab}(b)) 
contain armchair segments which mix the valleys and activate the backscattering. For the induced ring this is no longer the case.
In order to observe the AB oscillations in Fig.~\ref{zzt} we needed to introduce a short-range disorder to activate the intervalley scattering. 
The disorder is introduced at the edge, with a random removal of carbon atoms in a region of width 10 nm and length 540 nm, close to the upper ribbon edge (with probability of atom removal 0.01). The width of the zigzag ribbon both with and without defects is 97.42 nm.
The conductance for the ribbon with defects is plotted with the red line in Fig.~\ref{zzt} and exhibits the oscillations missing for the perfect zigzag edge.

\begin{figure}[tb!]
  \includegraphics[width= \columnwidth]{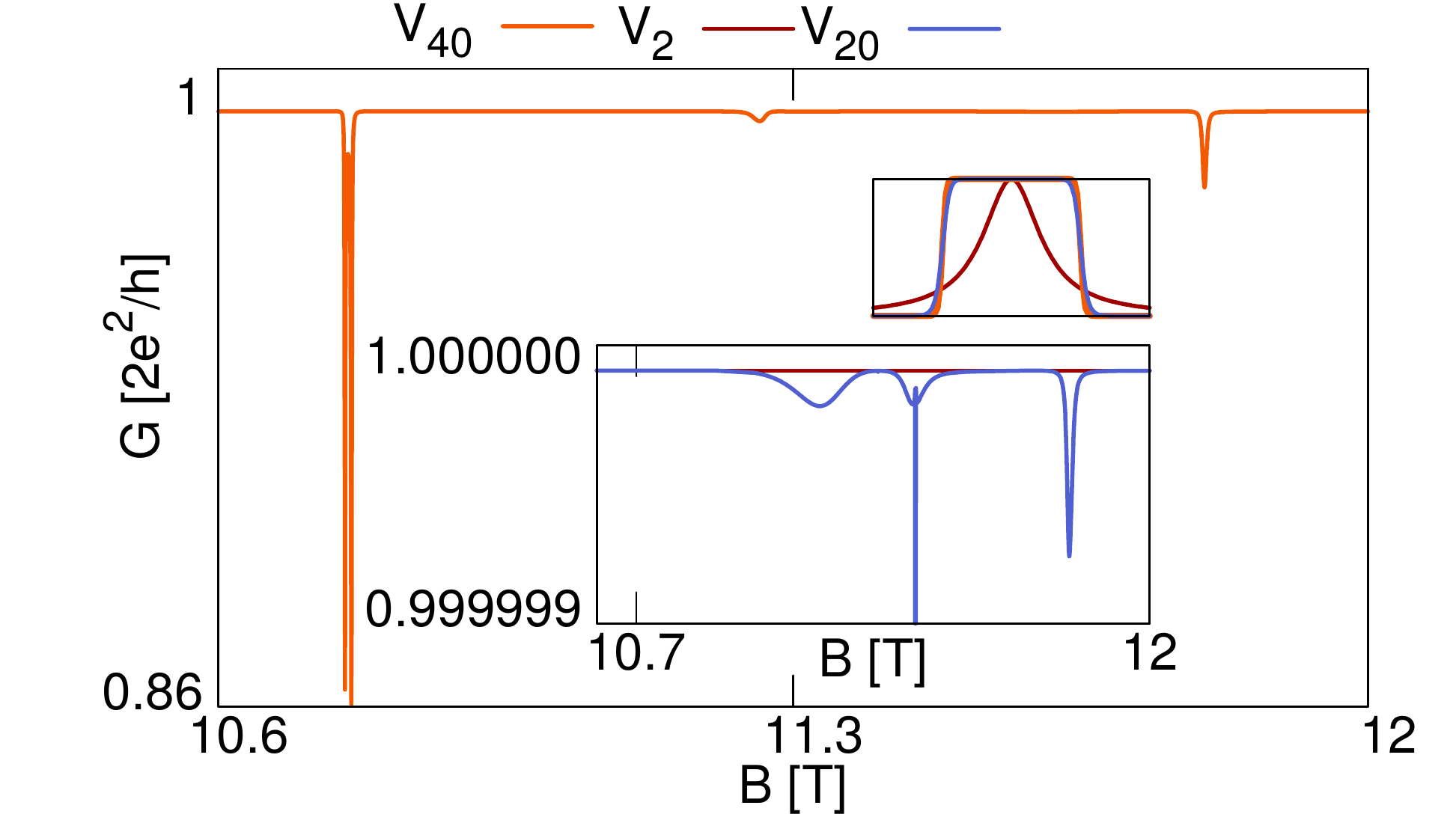}
  \caption{The conductance for the system if Fig. \ref{zzt} for the Lorentzian tip potential $V_2$ and for the exponent $n$ is \eqref{lf}
replaced by 20 for $V_{20}$ and 40 for $V_{40}$. The higher inset shows the profile of the tip potential and the lower one - a zoom of the main figure with $V_2$ and $V_{20}$ only. }\label{zztn}
\end{figure}

Without the defects and for the Lorentzian tip potential the quasibound states are valley degenerate (see Appendix A) and they do not introduce any intervalley mixing. 
In order to observe the intevalley scattering by the quasibound states the external potential needs to possess a component that is short range on the scale of the crystal constant.
We replaced the exponent $n=2$ in \eqref{lf} by 20 and 40 and obtained the results that are displayed in Fig. \ref{zztn}. For higher exponent the potential is steeper,
and for $n=40$ the variation of the potential in space induces a significant backscattering by the quasibound states that appear with an Aharonov-Bohm periodicity. 

\section{Summary and Conclusions}
We have studied the coherent transport through Aharonov-Bohm (AB) interferometers with magnetic deflection of electron currents. We considered both etched quantum rings and the ones induced by external potential. We solved  the quantum scattering problem and determined the quasibound states using an atomistic tight-binding approach. 

We identified two series of quasibound states with opposite orientation of  the currents.  The orientation
of the magnetic dipole and thus the energy shift
of the resonances and antiresonances was observed in conductance.
The magnetic forces pushing the currents to the internal or external edge depending on the orientation distinctly influence the AB oscillation period at the magnetic field scale.
Moreover, the Lorentz force increases or decreases the coupling of the ring-localized resonances with the leads depending on the orientation of the current. This amounts in modification  
of the lifetime of the quasibound states that in turn determines the width of conductance extrema that appear due to the interference of the incident electron wave function with the resonant localized states. Stabilization of the resonances with currents that produce the magnetic dipole that is antiparallel to the external field produces high harmonics of the conductance at high magnetic field.

\appendix 
\section{Stabilization method}
\renewcommand\thefigure{\thesection.\arabic{figure}}
\setcounter{figure}{0}
The stabilization method \cite{mandelsh} is used for detection of the resonant states localized within the ring. 
The number of eigenvalues is then normalized to 1, giving the fraction of energy levels in an energy range.
The method \cite{mandelsh} extracts the states that are localized within the ring as their
wave functions penetrate only weakly the leads and the dependence of the energy levels on the length of the leads is also weak.	
Throughout the paper, as a resonance counter we plot a fraction of energy levels $F$ found in a given energy range. 
 We solve the eigenproblem of the Hamiltonian (\ref{eq:dh}) for a closed system of the ring connected to finite size ribbons, which are extended  as marked in Fig. \ref{schemstab}. The energy spectrum is calculated as a function of the lengths of the leads.

\begin{figure}[tb!]
 \includegraphics[width=0.33\columnwidth]{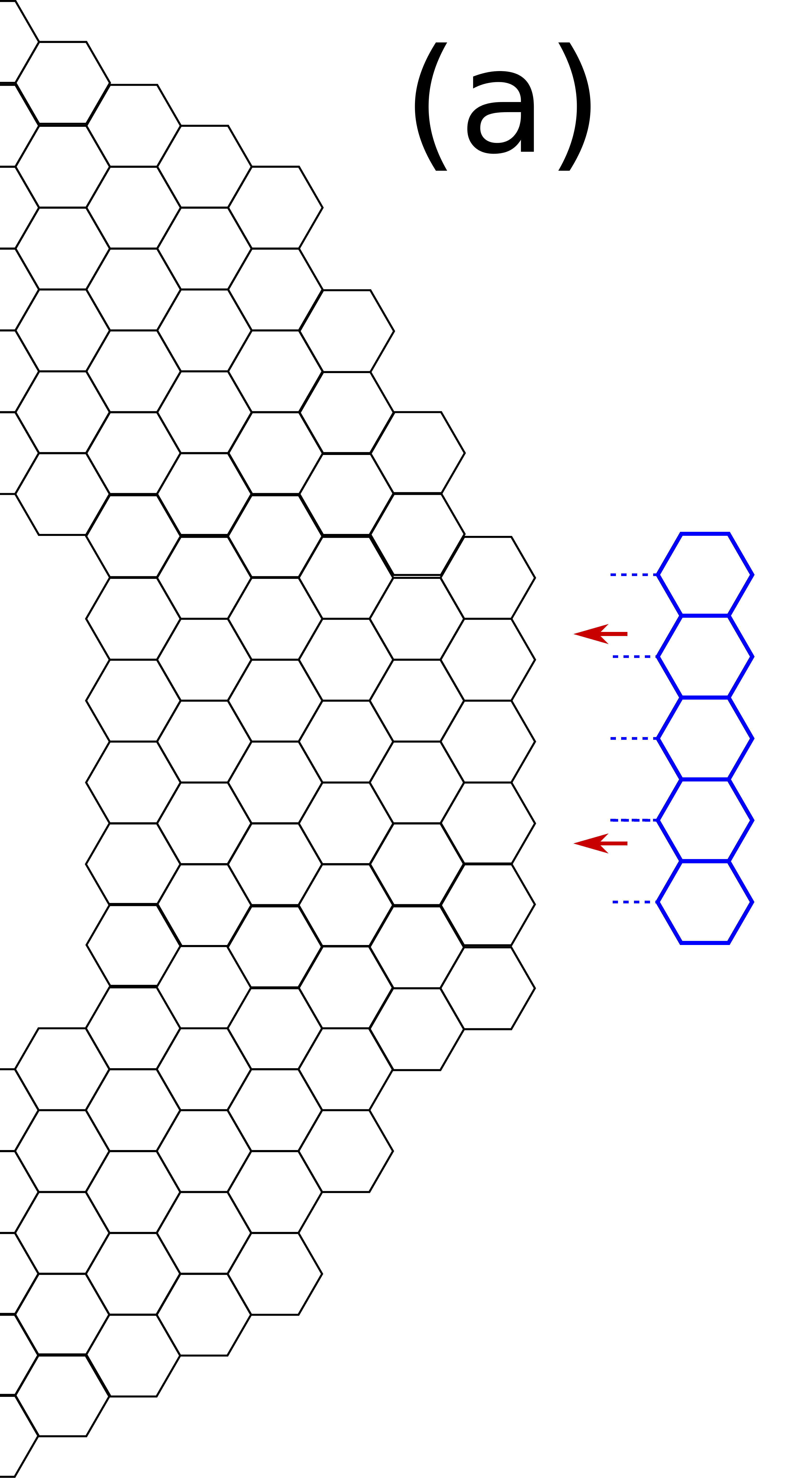}
 \includegraphics[width=0.29\columnwidth]{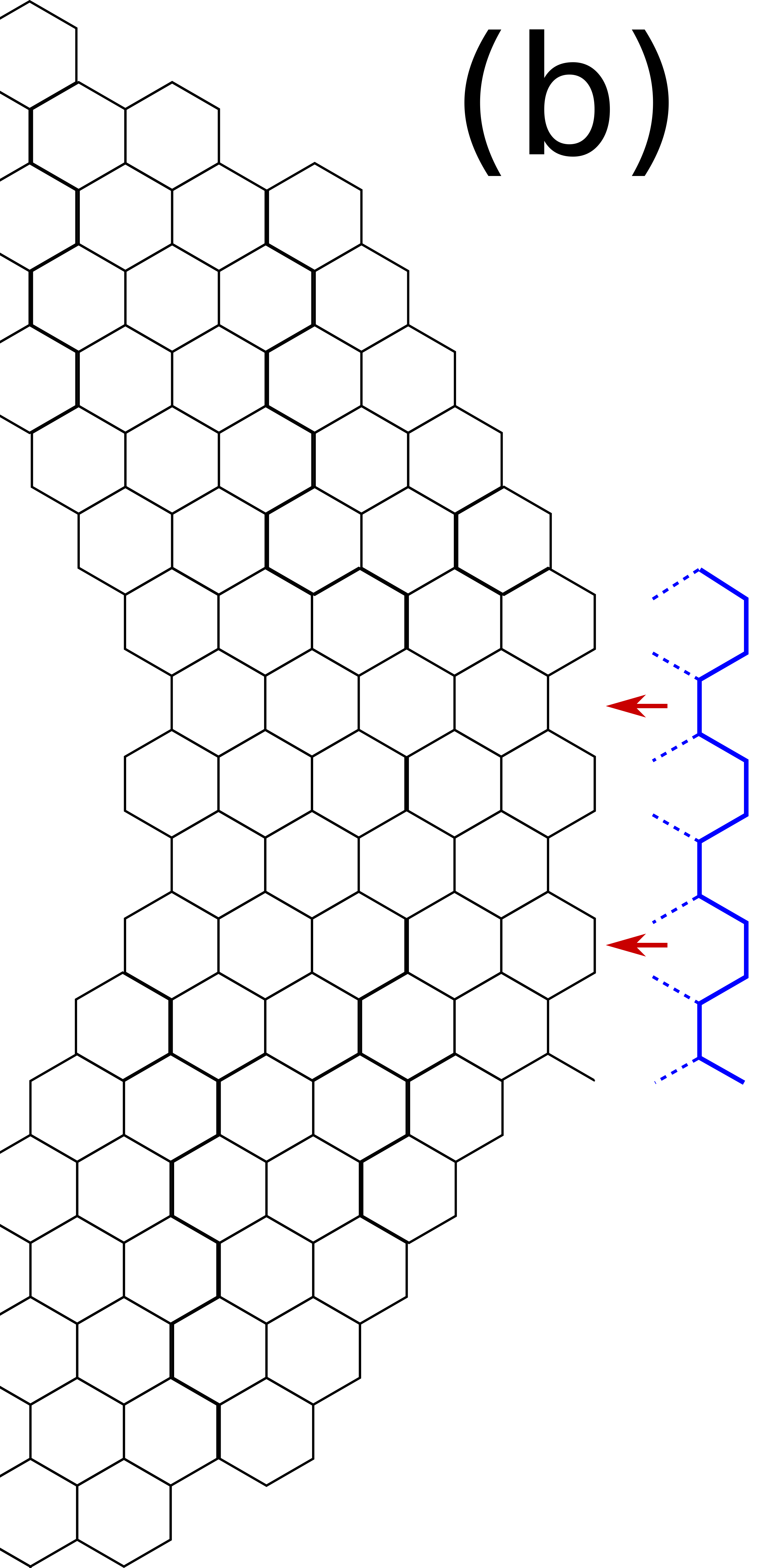}
  \caption{Scheme of the extending of the leads length for the calculation of the density of ring-localized resonances for an armchair (a) and zigzag (b) nanoribbons
serving as leads.
  } \label{schemstab}
\end{figure}

\begin{figure}[tb!]
 \includegraphics[width=\columnwidth]{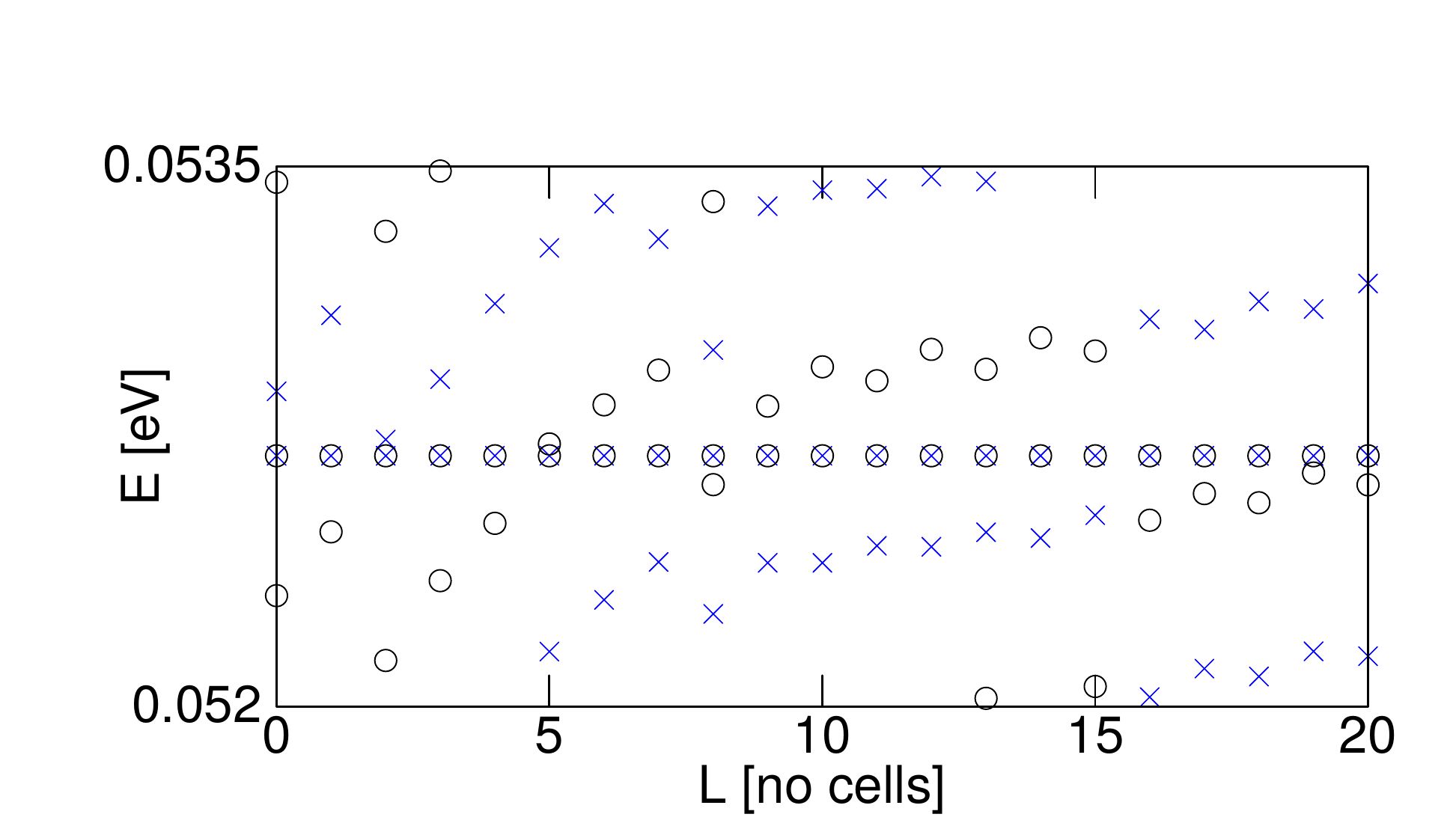}
  \caption{ Energy spectrum calculated as a function of the lengths of the leads of a zigzag nanoribbon with a Lorentzian potential discussed in the paper, for $B=6.6$ T. Every second energy level is marked with a cross or a circle. The energy levels that ar nearly independent of the leads length -- corresponding to a localized resonance -- are twofold degenerate. 
  } \label{EL}
\end{figure}

For the etched ring we start from the ring alone, then add subsequently single 
layers of carbon atoms at each side of the ring, each time evaluating the eigenvalues, until we reach 30 (20) layers for devices with armchair (zigzag) leads.
For the nanoribbon with tip, we begin with a ribbon 116 nm (148 nm) long for armchair (zigzag) ribbon, and proceed similarly as for the nanoring.

An example of the energy spectrum as a function of the leads length is shown in Fig. \ref{EL} for a zigzag nanoribbon with a Lorentzian potential as discussed in Section III.D, for $B=6.6$ T, in a narrow range of energy.

 Then, we count the eigenvalues in a small energy window obtained for varied length of the leads , using the detection counter \cite{mandelsh,Poniedzialek}:
\begin{equation}
   F(E)=\sum\limits_L \sum\limits_i \delta( |E-E_i(L)|;dE ), 
\label{eq:counter}
\end{equation}
where the first sum runs over the number of segments in the cell $L$, and the second sum over the eigenenergies $i$ of the closed system. Here, we define

\begin{equation}
  \delta( |E-E_i(L)|;dE) = \left\{ \begin{matrix}
  0; |E_i(L)-E|>dE \\
  1; |E_i(L)-E|\le dE
  \end{matrix} \right. .
\label{eq:}
\end{equation}

\section{Non-zero temperature and finite bias}
The results presented in the paper were obtained for zero temperature and in the linear response regime.
Here we evaluate the effects of a finite temperature and a finite bias for the conductance and its harmonics.
 For non-zero temperature, we calculate conductance by integration of conductance over the energy window \cite{datta}
\begin{equation}
 G( E_F;T ) = \frac{2e^2}{h} \int T_{tot}(E) \left( -\frac{\partial f_{E_F}(E) }{\partial E} \right) dE,
\end{equation}
defined by derivative of the  Fermi-Dirac distribution $f_{E_F}(E) = 1/( \exp( (E-E_F)/kT )+1 )$.

\begin{figure}[tb!]
 \includegraphics[width=\columnwidth]{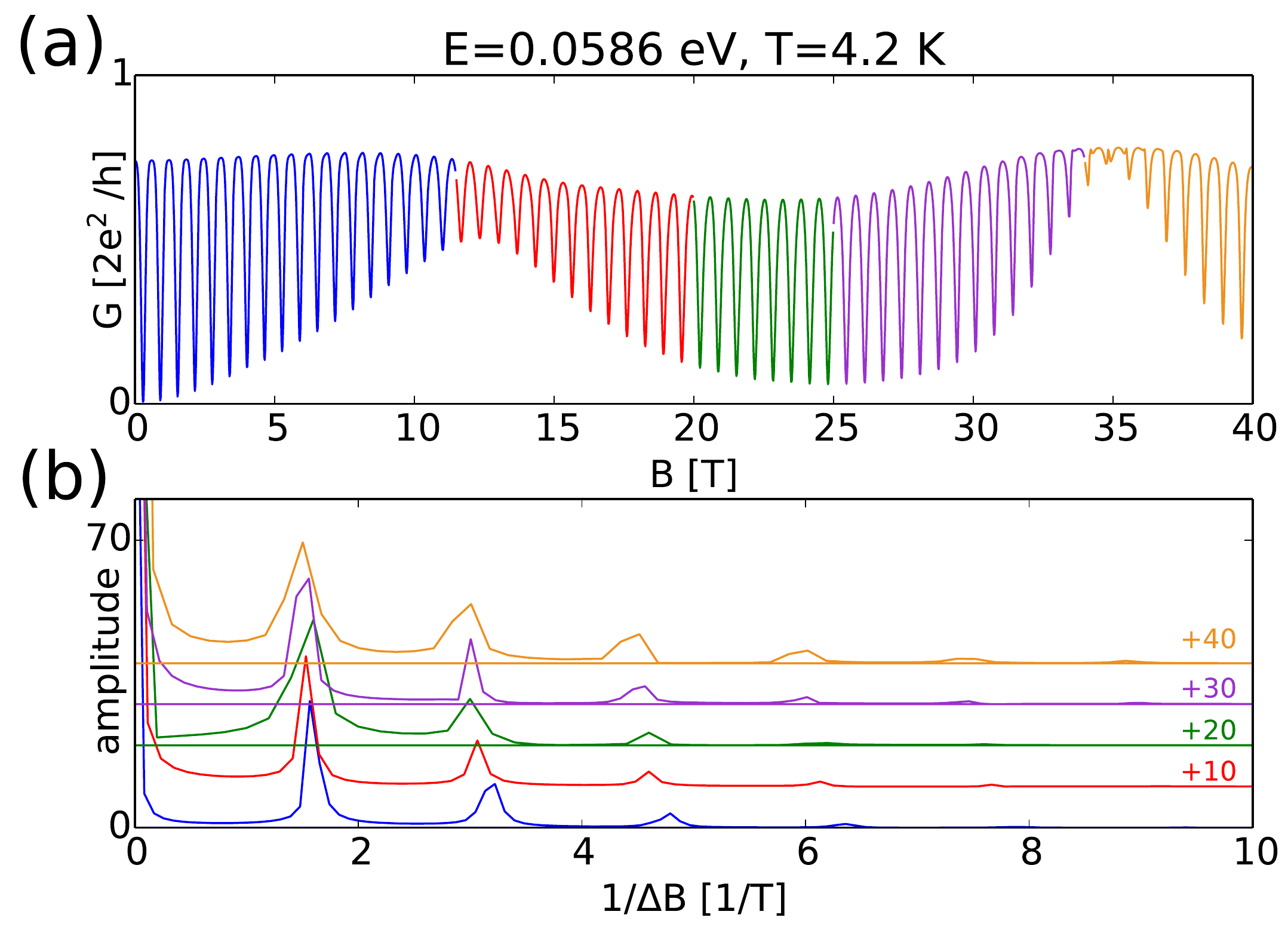}
  \caption{Same as Fig. \ref{fft1} only for $T= 4.2$K.}\label{40k}
\end{figure}

\begin{figure}[tb!]
 \includegraphics[width=\columnwidth]{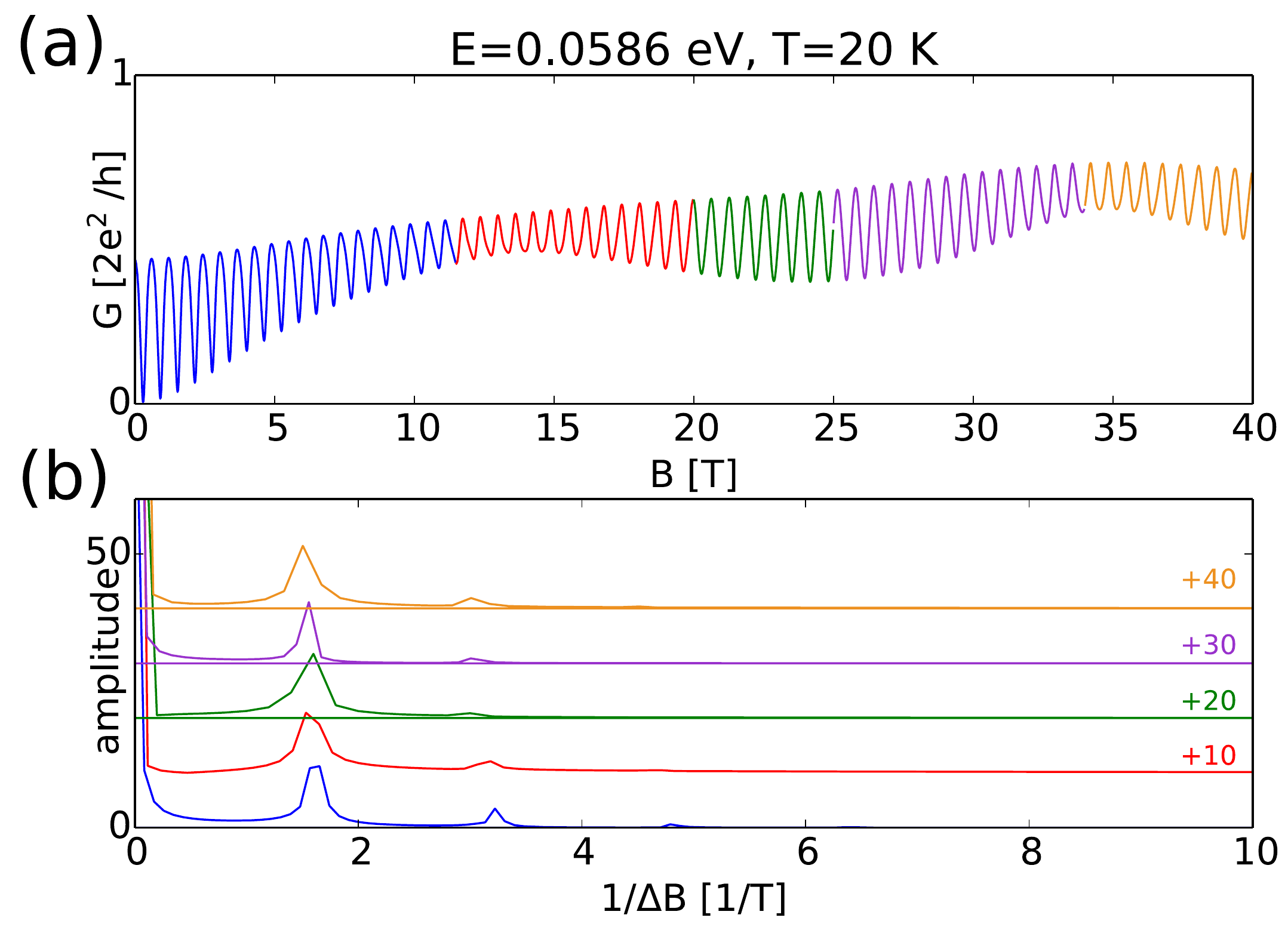}
  \caption{Same as Fig. \ref{fft1} only for $T= 20$K.}\label{20k}
\end{figure}

Figures   \ref{40k} (for $T=4.2$ K)  and \ref{20k} (for $T=20$K) as compared to Fig. \ref{fft1} (for $T=0$) demonstrate that with a growing temperature the amplitude 
of the conductance oscillations is decreased and the contribution of the higher harmonics are decreased.

For conductance in a finite source-drain bias, applied to cover the non-linear transport regime, we apply voltages $+ \tfrac{V_{SD}}{2}$ $\left(- \tfrac{V_{SD}}{2} \right)$ in the left (right) lead, and 
assume a linear change of the voltage along the system. %: 
%\begin{equation}
% V( x,y ) = -\frac{V_{SD}}{2 L_D}  \left( x- \frac{L_D }{2} \right),
%\end{equation}
%where $L_D$ is the length of the system.
We calculate the current with the formula \cite{datta}
\begin{equation}
 I( E_F;T=0, V_{SD} ) = \frac{e}{h} \int\limits_{-eV_{SD}/2}^{+eV_{SD}/2} T_{tot}(E_F+E) dE.
\end{equation}

\begin{figure}[tb!]
 \includegraphics[width=\columnwidth]{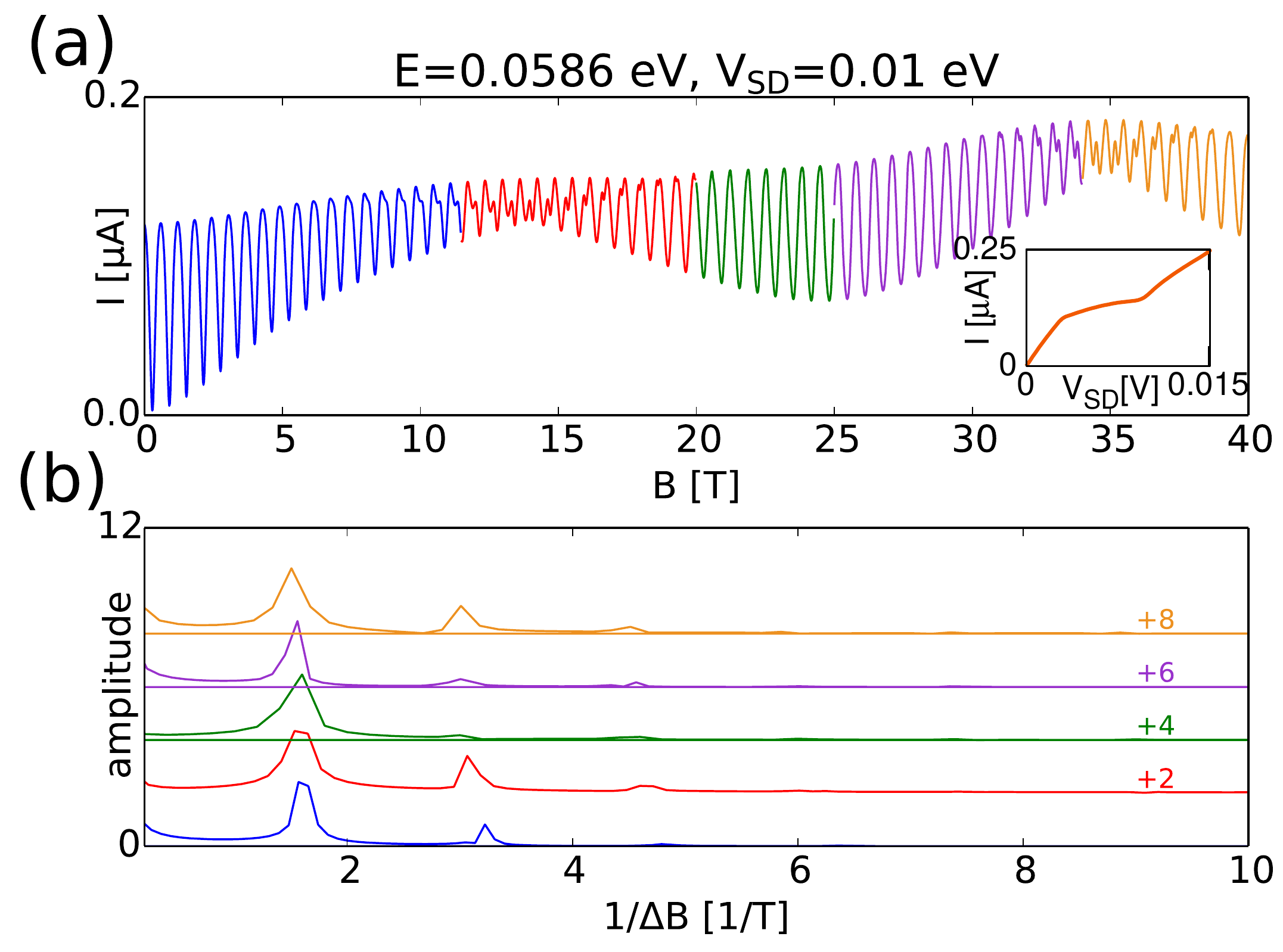}
  \caption{Same as Fig. \ref{fft1} only for current at source drain bias of 10 meV. The inset to (a) shows the current - bias characteristics.}\label{pro}
\end{figure}

The linear conductance regime -- see the inset to Fig. \ref{pro} covers the bias up to 15 meV.
We considered $V_{sd}=10$ meV -- outside the linear regime and calculated the conductance dependence on the external 
magnetic field. The effect of a finite bias for the conductance that we find here is similar to the one obtained
for a finite temperature: the visibility of the conductance oscillations are reduced and the higher harmonics are attenuated.
For both the finite temperature and the bias we deal with a finite energy window for the transport. 
In the nonlinear transport conditions the effects of the electron-electron interaction \cite{Anjou,Kubo,Bedkihal} may lead to $dI/dB|_{B=0}\neq 0 $
and to an even-odd effects for conductance harmonics \cite{Anjou}. The present modelling neglects the interaction
and hence these effects are not observed.
 
\section*{Acknowledgments}
This work was supported by the National Science Centre (NCN) according to decision DEC-2015/17/B/ST3/01161.
A. M.-K. acknowledges the doctoral stipend
Etiuda funded by the National Science Centre (NCN) according to DEC-2015/16/T/ST3/00264.
The calculations were performed on PL-Grid Infrastructure.

\bibliography{RingBib}

\begin{thebibliography}{59}%
\makeatletter
\providecommand \@ifxundefined [1]{%
 \@ifx{#1\undefined}
}%
\providecommand \@ifnum [1]{%
 \ifnum #1\expandafter \@firstoftwo
 \else \expandafter \@secondoftwo
 \fi
}%
\providecommand \@ifx [1]{%
 \ifx #1\expandafter \@firstoftwo
 \else \expandafter \@secondoftwo
 \fi
}%
\providecommand \natexlab [1]{#1}%
\providecommand \enquote  [1]{``#1''}%
\providecommand \bibnamefont  [1]{#1}%
\providecommand \bibfnamefont [1]{#1}%
\providecommand \citenamefont [1]{#1}%
\providecommand \href@noop [0]{\@secondoftwo}%
\providecommand \href [0]{\begingroup \@sanitize@url \@href}%
\providecommand \@href[1]{\@@startlink{#1}\@@href}%
\providecommand \@@href[1]{\endgroup#1\@@endlink}%
\providecommand \@sanitize@url [0]{\catcode `\\12\catcode `\$12\catcode
  `\&12\catcode `\#12\catcode `\^12\catcode `\_12\catcode `\%12\relax}%
\providecommand \@@startlink[1]{}%
\providecommand \@@endlink[0]{}%
\providecommand \url  [0]{\begingroup\@sanitize@url \@url }%
\providecommand \@url [1]{\endgroup\@href {#1}{\urlprefix }}%
\providecommand \urlprefix  [0]{URL }%
\providecommand \Eprint [0]{\href }%
\providecommand \doibase [0]{http://dx.doi.org/}%
\providecommand \selectlanguage [0]{\@gobble}%
\providecommand \bibinfo  [0]{\@secondoftwo}%
\providecommand \bibfield  [0]{\@secondoftwo}%
\providecommand \translation [1]{[#1]}%
\providecommand \BibitemOpen [0]{}%
\providecommand \bibitemStop [0]{}%
\providecommand \bibitemNoStop [0]{.\EOS\space}%
\providecommand \EOS [0]{\spacefactor3000\relax}%
\providecommand \BibitemShut  [1]{\csname bibitem#1\endcsname}%
\let\auto@bib@innerbib\@empty
%</preamble>
\bibitem [{\citenamefont {Aronov}\ and\ \citenamefont
  {Sharvin}(1987)}]{Aronov87}%
  \BibitemOpen
  \bibfield  {author} {\bibinfo {author} {\bibfnamefont {A.~G.}\ \bibnamefont
  {Aronov}}\ and\ \bibinfo {author} {\bibfnamefont {Y.~V.}\ \bibnamefont
  {Sharvin}},\ }\href {\doibase 10.1103/RevModPhys.59.755} {\bibfield
  {journal} {\bibinfo  {journal} {Rev. Mod. Phys.}\ }\textbf {\bibinfo {volume}
  {59}},\ \bibinfo {pages} {755} (\bibinfo {year} {1987})}\BibitemShut
  {NoStop}%
\bibitem [{\citenamefont {B\"uttiker}\ \emph {et~al.}(1983)\citenamefont
  {B\"uttiker}, \citenamefont {Imry},\ and\ \citenamefont
  {Landauer}}]{Buttiker83}%
  \BibitemOpen
  \bibfield  {author} {\bibinfo {author} {\bibfnamefont {M.}~\bibnamefont
  {B\"uttiker}}, \bibinfo {author} {\bibfnamefont {Y.}~\bibnamefont {Imry}}, \
  and\ \bibinfo {author} {\bibfnamefont {R.}~\bibnamefont {Landauer}},\
  }\href@noop {} {\bibfield  {journal} {\bibinfo  {journal} {Phys. Lett. A}\
  }\textbf {\bibinfo {volume} {96}},\ \bibinfo {pages} {365 } (\bibinfo {year}
  {1983})}\BibitemShut {NoStop}%
\bibitem [{\citenamefont {Webb}\ \emph {et~al.}(1985)\citenamefont {Webb},
  \citenamefont {Washburn}, \citenamefont {Umbach},\ and\ \citenamefont
  {Laibowitz}}]{Webb85}%
  \BibitemOpen
  \bibfield  {author} {\bibinfo {author} {\bibfnamefont {R.~A.}\ \bibnamefont
  {Webb}}, \bibinfo {author} {\bibfnamefont {S.}~\bibnamefont {Washburn}},
  \bibinfo {author} {\bibfnamefont {C.~P.}\ \bibnamefont {Umbach}}, \ and\
  \bibinfo {author} {\bibfnamefont {R.~B.}\ \bibnamefont {Laibowitz}},\
  }\href@noop {} {\bibfield  {journal} {\bibinfo  {journal} {Phys. Rev. Lett.}\
  }\textbf {\bibinfo {volume} {54}},\ \bibinfo {pages} {2696} (\bibinfo {year}
  {1985})}\BibitemShut {NoStop}%
\bibitem [{\citenamefont {Timp}\ \emph {et~al.}(1987)\citenamefont {Timp},
  \citenamefont {Chang}, \citenamefont {Cunningham}, \citenamefont {Chang},
  \citenamefont {Mankiewich}, \citenamefont {Behringer},\ and\ \citenamefont
  {Howard}}]{Timp87}%
  \BibitemOpen
  \bibfield  {author} {\bibinfo {author} {\bibfnamefont {G.}~\bibnamefont
  {Timp}}, \bibinfo {author} {\bibfnamefont {A.~M.}\ \bibnamefont {Chang}},
  \bibinfo {author} {\bibfnamefont {J.~E.}\ \bibnamefont {Cunningham}},
  \bibinfo {author} {\bibfnamefont {T.~Y.}\ \bibnamefont {Chang}}, \bibinfo
  {author} {\bibfnamefont {P.}~\bibnamefont {Mankiewich}}, \bibinfo {author}
  {\bibfnamefont {R.}~\bibnamefont {Behringer}}, \ and\ \bibinfo {author}
  {\bibfnamefont {R.~E.}\ \bibnamefont {Howard}},\ }\href@noop {} {\bibfield
  {journal} {\bibinfo  {journal} {Phys. Rev. Lett.}\ }\textbf {\bibinfo
  {volume} {58}},\ \bibinfo {pages} {2814} (\bibinfo {year}
  {1987})}\BibitemShut {NoStop}%
\bibitem [{\citenamefont {Kvon}\ \emph {et~al.}(2008)\citenamefont {Kvon},
  \citenamefont {Kozlov}, \citenamefont {Olshanetsky}, \citenamefont
  {Plotnikov}, \citenamefont {Latyshev},\ and\ \citenamefont {Portal}}]{kvon}%
  \BibitemOpen
  \bibfield  {author} {\bibinfo {author} {\bibfnamefont {Z.~D.}\ \bibnamefont
  {Kvon}}, \bibinfo {author} {\bibfnamefont {D.~A.}\ \bibnamefont {Kozlov}},
  \bibinfo {author} {\bibfnamefont {E.~B.}\ \bibnamefont {Olshanetsky}},
  \bibinfo {author} {\bibfnamefont {A.~E.}\ \bibnamefont {Plotnikov}}, \bibinfo
  {author} {\bibfnamefont {A.~V.}\ \bibnamefont {Latyshev}}, \ and\ \bibinfo
  {author} {\bibfnamefont {J.~C.}\ \bibnamefont {Portal}},\ }\href {\doibase
  http://dx.doi.org/10.1016/j.ssc.2008.05.019} {\bibfield  {journal} {\bibinfo
  {journal} {Sol. Stat. Comm.}\ }\textbf {\bibinfo {volume} {147}},\ \bibinfo
  {pages} {230 } (\bibinfo {year} {2008})}\BibitemShut {NoStop}%
\bibitem [{\citenamefont {Aharonov}\ and\ \citenamefont {Bohm}(1959)}]{AB}%
  \BibitemOpen
  \bibfield  {author} {\bibinfo {author} {\bibfnamefont {Y.}~\bibnamefont
  {Aharonov}}\ and\ \bibinfo {author} {\bibfnamefont {D.}~\bibnamefont
  {Bohm}},\ }\href@noop {} {\bibfield  {journal} {\bibinfo  {journal} {Phys.
  Rev.}\ }\textbf {\bibinfo {volume} {115}},\ \bibinfo {pages} {485} (\bibinfo
  {year} {1959})}\BibitemShut {NoStop}%
\bibitem [{\citenamefont {Szafran}\ and\ \citenamefont
  {Peeters}(2005{\natexlab{a}})}]{Sza0}%
  \BibitemOpen
  \bibfield  {author} {\bibinfo {author} {\bibfnamefont {B.}~\bibnamefont
  {Szafran}}\ and\ \bibinfo {author} {\bibfnamefont {F.~M.}\ \bibnamefont
  {Peeters}},\ }\href@noop {} {\bibfield  {journal} {\bibinfo  {journal} {Phys.
  Rev. B}\ }\textbf {\bibinfo {volume} {72}},\ \bibinfo {pages} {165301}
  (\bibinfo {year} {2005}{\natexlab{a}})}\BibitemShut {NoStop}%
\bibitem [{\citenamefont {Szafran}\ and\ \citenamefont
  {Peeters}(2005{\natexlab{b}})}]{Sza1}%
  \BibitemOpen
  \bibfield  {author} {\bibinfo {author} {\bibfnamefont {B.}~\bibnamefont
  {Szafran}}\ and\ \bibinfo {author} {\bibfnamefont {F.~M.}\ \bibnamefont
  {Peeters}},\ }\href@noop {} {\bibfield  {journal} {\bibinfo  {journal} {Euro.
  Lett.}\ }\textbf {\bibinfo {volume} {70}},\ \bibinfo {pages} {810} (\bibinfo
  {year} {2005}{\natexlab{b}})}\BibitemShut {NoStop}%
\bibitem [{\citenamefont {Aidala}\ \emph {et~al.}(2007)\citenamefont {Aidala},
  \citenamefont {Parrott}, \citenamefont {Kramer}, \citenamefont {Heller},
  \citenamefont {Westervelt}, \citenamefont {Hanson},\ and\ \citenamefont
  {Gossard}}]{MF0}%
  \BibitemOpen
  \bibfield  {author} {\bibinfo {author} {\bibfnamefont {K.~E.}\ \bibnamefont
  {Aidala}}, \bibinfo {author} {\bibfnamefont {R.~E.}\ \bibnamefont {Parrott}},
  \bibinfo {author} {\bibfnamefont {T.}~\bibnamefont {Kramer}}, \bibinfo
  {author} {\bibfnamefont {E.~J.}\ \bibnamefont {Heller}}, \bibinfo {author}
  {\bibfnamefont {R.~M.}\ \bibnamefont {Westervelt}}, \bibinfo {author}
  {\bibfnamefont {M.~P.}\ \bibnamefont {Hanson}}, \ and\ \bibinfo {author}
  {\bibfnamefont {A.~C.}\ \bibnamefont {Gossard}},\ }\href@noop {} {\bibfield
  {journal} {\bibinfo  {journal} {Nat. Phys.}\ }\textbf {\bibinfo {volume}
  {3}},\ \bibinfo {pages} {464} (\bibinfo {year} {2007})}\BibitemShut {NoStop}%
\bibitem [{\citenamefont {Taychatanapat}\ \emph {et~al.}(2013)\citenamefont
  {Taychatanapat}, \citenamefont {Watanabe}, \citenamefont {Taniguchi},\ and\
  \citenamefont {Jarillo-Herrero}}]{MF1}%
  \BibitemOpen
  \bibfield  {author} {\bibinfo {author} {\bibfnamefont {T.}~\bibnamefont
  {Taychatanapat}}, \bibinfo {author} {\bibfnamefont {K.}~\bibnamefont
  {Watanabe}}, \bibinfo {author} {\bibfnamefont {T.}~\bibnamefont {Taniguchi}},
  \ and\ \bibinfo {author} {\bibfnamefont {P.}~\bibnamefont
  {Jarillo-Herrero}},\ }\href@noop {} {\bibfield  {journal} {\bibinfo
  {journal} {Nat. Phys.}\ }\textbf {\bibinfo {volume} {9}},\ \bibinfo {pages}
  {225} (\bibinfo {year} {2013})}\BibitemShut {NoStop}%
\bibitem [{\citenamefont {Bhandari}\ \emph {et~al.}(2016)\citenamefont
  {Bhandari}, \citenamefont {Lee}, \citenamefont {Klales}, \citenamefont
  {Watanabe}, \citenamefont {Taniguchi}, \citenamefont {Heller}, \citenamefont
  {Kim},\ and\ \citenamefont {Westervelt}}]{MF2}%
  \BibitemOpen
  \bibfield  {author} {\bibinfo {author} {\bibfnamefont {S.}~\bibnamefont
  {Bhandari}}, \bibinfo {author} {\bibfnamefont {G.-H.}\ \bibnamefont {Lee}},
  \bibinfo {author} {\bibfnamefont {A.}~\bibnamefont {Klales}}, \bibinfo
  {author} {\bibfnamefont {K.}~\bibnamefont {Watanabe}}, \bibinfo {author}
  {\bibfnamefont {T.}~\bibnamefont {Taniguchi}}, \bibinfo {author}
  {\bibfnamefont {E.}~\bibnamefont {Heller}}, \bibinfo {author} {\bibfnamefont
  {P.}~\bibnamefont {Kim}}, \ and\ \bibinfo {author} {\bibfnamefont {R.~M.}\
  \bibnamefont {Westervelt}},\ }\href@noop {} {\bibfield  {journal} {\bibinfo
  {journal} {Nano Lett.}\ }\textbf {\bibinfo {volume} {16}},\ \bibinfo {pages}
  {1690} (\bibinfo {year} {2016})}\BibitemShut {NoStop}%
\bibitem [{\citenamefont {Chen}\ \emph {et~al.}(2013)\citenamefont {Chen},
  \citenamefont {Pepper}, \citenamefont {Farrer}, \citenamefont {Ritchie},\
  and\ \citenamefont {Jones}}]{MF3}%
  \BibitemOpen
  \bibfield  {author} {\bibinfo {author} {\bibfnamefont {T.-M.}\ \bibnamefont
  {Chen}}, \bibinfo {author} {\bibfnamefont {M.}~\bibnamefont {Pepper}},
  \bibinfo {author} {\bibfnamefont {I.}~\bibnamefont {Farrer}}, \bibinfo
  {author} {\bibfnamefont {D.~A.}\ \bibnamefont {Ritchie}}, \ and\ \bibinfo
  {author} {\bibfnamefont {G.~A.~C.}\ \bibnamefont {Jones}},\ }\href
  {http://scitation.aip.org/content/aip/journal/apl/103/9/10.1063/1.4819489}
  {\bibfield  {journal} {\bibinfo  {journal} {Appl. Phys. Lett.}\ }\textbf
  {\bibinfo {volume} {103}},\ \bibinfo {eid} {093503} (\bibinfo {year}
  {2013})}\BibitemShut {NoStop}%
\bibitem [{\citenamefont {Morikawa}\ \emph {et~al.}(2015)\citenamefont
  {Morikawa}, \citenamefont {Dou}, \citenamefont {Wang}, \citenamefont {Smith},
  \citenamefont {Watanabe}, \citenamefont {Taniguchi}, \citenamefont
  {Masubuchi}, \citenamefont {Machida},\ and\ \citenamefont {Connolly}}]{MF4}%
  \BibitemOpen
  \bibfield  {author} {\bibinfo {author} {\bibfnamefont {S.}~\bibnamefont
  {Morikawa}}, \bibinfo {author} {\bibfnamefont {Z.}~\bibnamefont {Dou}},
  \bibinfo {author} {\bibfnamefont {S.-W.}\ \bibnamefont {Wang}}, \bibinfo
  {author} {\bibfnamefont {C.~G.}\ \bibnamefont {Smith}}, \bibinfo {author}
  {\bibfnamefont {K.}~\bibnamefont {Watanabe}}, \bibinfo {author}
  {\bibfnamefont {T.}~\bibnamefont {Taniguchi}}, \bibinfo {author}
  {\bibfnamefont {S.}~\bibnamefont {Masubuchi}}, \bibinfo {author}
  {\bibfnamefont {T.}~\bibnamefont {Machida}}, \ and\ \bibinfo {author}
  {\bibfnamefont {M.~R.}\ \bibnamefont {Connolly}},\ }\href
  {http://scitation.aip.org/content/aip/journal/apl/107/24/10.1063/1.4937473}
  {\bibfield  {journal} {\bibinfo  {journal} {Appl. Phys. Lett.}\ }\textbf
  {\bibinfo {volume} {107}},\ \bibinfo {eid} {243102} (\bibinfo {year}
  {2015})}\BibitemShut {NoStop}%
\bibitem [{\citenamefont {Abanin}\ \emph {et~al.}(2007)\citenamefont {Abanin},
  \citenamefont {Novoselov}, \citenamefont {Zeitler}, \citenamefont {Lee},
  \citenamefont {Geim},\ and\ \citenamefont {Levitov}}]{DAbanin}%
  \BibitemOpen
  \bibfield  {author} {\bibinfo {author} {\bibfnamefont {D.~A.}\ \bibnamefont
  {Abanin}}, \bibinfo {author} {\bibfnamefont {K.~S.}\ \bibnamefont
  {Novoselov}}, \bibinfo {author} {\bibfnamefont {U.}~\bibnamefont {Zeitler}},
  \bibinfo {author} {\bibfnamefont {P.~A.}\ \bibnamefont {Lee}}, \bibinfo
  {author} {\bibfnamefont {A.~K.}\ \bibnamefont {Geim}}, \ and\ \bibinfo
  {author} {\bibfnamefont {L.~S.}\ \bibnamefont {Levitov}},\ }\href {\doibase
  10.1103/PhysRevLett.98.196806} {\bibfield  {journal} {\bibinfo  {journal}
  {Phys. Rev. Lett.}\ }\textbf {\bibinfo {volume} {98}},\ \bibinfo {pages}
  {196806} (\bibinfo {year} {2007})}\BibitemShut {NoStop}%
\bibitem [{\citenamefont {Cabosart}\ \emph {et~al.}(2014)\citenamefont
  {Cabosart}, \citenamefont {Faniel}, \citenamefont {Martins}, \citenamefont
  {Brun}, \citenamefont {Felten}, \citenamefont {Bayot},\ and\ \citenamefont
  {Hackens}}]{eg1}%
  \BibitemOpen
  \bibfield  {author} {\bibinfo {author} {\bibfnamefont {D.}~\bibnamefont
  {Cabosart}}, \bibinfo {author} {\bibfnamefont {S.}~\bibnamefont {Faniel}},
  \bibinfo {author} {\bibfnamefont {F.}~\bibnamefont {Martins}}, \bibinfo
  {author} {\bibfnamefont {B.}~\bibnamefont {Brun}}, \bibinfo {author}
  {\bibfnamefont {A.}~\bibnamefont {Felten}}, \bibinfo {author} {\bibfnamefont
  {V.}~\bibnamefont {Bayot}}, \ and\ \bibinfo {author} {\bibfnamefont
  {B.}~\bibnamefont {Hackens}},\ }\href@noop {} {\bibfield  {journal} {\bibinfo
   {journal} {Phys. Rev. B}\ }\textbf {\bibinfo {volume} {90}},\ \bibinfo
  {pages} {205433} (\bibinfo {year} {2014})}\BibitemShut {NoStop}%
\bibitem [{\citenamefont {Russo}\ \emph {et~al.}(2008)\citenamefont {Russo},
  \citenamefont {Oostinga}, \citenamefont {Wehenkel}, \citenamefont {Heersche},
  \citenamefont {Sobhani}, \citenamefont {Vandersypen},\ and\ \citenamefont
  {Morpurgo}}]{eg2}%
  \BibitemOpen
  \bibfield  {author} {\bibinfo {author} {\bibfnamefont {S.}~\bibnamefont
  {Russo}}, \bibinfo {author} {\bibfnamefont {J.~B.}\ \bibnamefont {Oostinga}},
  \bibinfo {author} {\bibfnamefont {D.}~\bibnamefont {Wehenkel}}, \bibinfo
  {author} {\bibfnamefont {H.~B.}\ \bibnamefont {Heersche}}, \bibinfo {author}
  {\bibfnamefont {S.~S.}\ \bibnamefont {Sobhani}}, \bibinfo {author}
  {\bibfnamefont {L.~M.~K.}\ \bibnamefont {Vandersypen}}, \ and\ \bibinfo
  {author} {\bibfnamefont {A.~F.}\ \bibnamefont {Morpurgo}},\ }\href@noop {}
  {\bibfield  {journal} {\bibinfo  {journal} {Phys. Rev. B}\ }\textbf {\bibinfo
  {volume} {77}},\ \bibinfo {pages} {085413} (\bibinfo {year}
  {2008})}\BibitemShut {NoStop}%
\bibitem [{\citenamefont {Smirnov}\ \emph {et~al.}(2012)\citenamefont
  {Smirnov}, \citenamefont {Schmidt},\ and\ \citenamefont {Haug}}]{eg3}%
  \BibitemOpen
  \bibfield  {author} {\bibinfo {author} {\bibfnamefont {D.}~\bibnamefont
  {Smirnov}}, \bibinfo {author} {\bibfnamefont {H.}~\bibnamefont {Schmidt}}, \
  and\ \bibinfo {author} {\bibfnamefont {R.~J.}\ \bibnamefont {Haug}},\ }\href
  {http://scitation.aip.org/content/aip/journal/apl/100/20/10.1063/1.4717622}
  {\bibfield  {journal} {\bibinfo  {journal} {Appl. Phys. Lett.}\ }\textbf
  {\bibinfo {volume} {100}},\ \bibinfo {eid} {203114} (\bibinfo {year}
  {2012})}\BibitemShut {NoStop}%
\bibitem [{\citenamefont {Smirnov}\ \emph {et~al.}(2014)\citenamefont
  {Smirnov}, \citenamefont {Rode},\ and\ \citenamefont {Haug}}]{eg4}%
  \BibitemOpen
  \bibfield  {author} {\bibinfo {author} {\bibfnamefont {D.}~\bibnamefont
  {Smirnov}}, \bibinfo {author} {\bibfnamefont {J.~C.}\ \bibnamefont {Rode}}, \
  and\ \bibinfo {author} {\bibfnamefont {R.~J.}\ \bibnamefont {Haug}},\ }\href
  {http://scitation.aip.org/content/aip/journal/apl/105/8/10.1063/1.4894471}
  {\bibfield  {journal} {\bibinfo  {journal} {Appl. Phys. Lett.}\ }\textbf
  {\bibinfo {volume} {105}},\ \bibinfo {eid} {082112} (\bibinfo {year}
  {2014})}\BibitemShut {NoStop}%
\bibitem [{\citenamefont {Huefner}\ \emph {et~al.}(2010)\citenamefont
  {Huefner}, \citenamefont {Molitor}, \citenamefont {Jacobsen}, \citenamefont
  {Pioda}, \citenamefont {Stampfer}, \citenamefont {Ensslin},\ and\
  \citenamefont {Ihn}}]{eg5}%
  \BibitemOpen
  \bibfield  {author} {\bibinfo {author} {\bibfnamefont {M.}~\bibnamefont
  {Huefner}}, \bibinfo {author} {\bibfnamefont {F.}~\bibnamefont {Molitor}},
  \bibinfo {author} {\bibfnamefont {A.}~\bibnamefont {Jacobsen}}, \bibinfo
  {author} {\bibfnamefont {A.}~\bibnamefont {Pioda}}, \bibinfo {author}
  {\bibfnamefont {C.}~\bibnamefont {Stampfer}}, \bibinfo {author}
  {\bibfnamefont {K.}~\bibnamefont {Ensslin}}, \ and\ \bibinfo {author}
  {\bibfnamefont {T.}~\bibnamefont {Ihn}},\ }\href@noop {} {\bibfield
  {journal} {\bibinfo  {journal} {New J. Phys.}\ }\textbf {\bibinfo {volume}
  {12}},\ \bibinfo {pages} {043054} (\bibinfo {year} {2010})}\BibitemShut
  {NoStop}%
\bibitem [{\citenamefont {Rahman}\ \emph {et~al.}(2013)\citenamefont {Rahman},
  \citenamefont {Guikema}, \citenamefont {Lee},\ and\ \citenamefont
  {Markovi\ifmmode~\acute{c}\else \'{c}\fi{}}}]{eg6}%
  \BibitemOpen
  \bibfield  {author} {\bibinfo {author} {\bibfnamefont {A.}~\bibnamefont
  {Rahman}}, \bibinfo {author} {\bibfnamefont {J.~W.}\ \bibnamefont {Guikema}},
  \bibinfo {author} {\bibfnamefont {S.~H.}\ \bibnamefont {Lee}}, \ and\
  \bibinfo {author} {\bibfnamefont {N.}~\bibnamefont
  {Markovi\ifmmode~\acute{c}\else \'{c}\fi{}}},\ }\href {\doibase
  10.1103/PhysRevB.87.081401} {\bibfield  {journal} {\bibinfo  {journal} {Phys.
  Rev. B}\ }\textbf {\bibinfo {volume} {87}},\ \bibinfo {pages} {081401(R)}
  (\bibinfo {year} {2013})}\BibitemShut {NoStop}%
\bibitem [{\citenamefont {Recher}\ \emph {et~al.}(2007)\citenamefont {Recher},
  \citenamefont {Trauzettel}, \citenamefont {Rycerz}, \citenamefont {Blanter},
  \citenamefont {Beenakker},\ and\ \citenamefont {Morpurgo}}]{tg1}%
  \BibitemOpen
  \bibfield  {author} {\bibinfo {author} {\bibfnamefont {P.}~\bibnamefont
  {Recher}}, \bibinfo {author} {\bibfnamefont {B.}~\bibnamefont {Trauzettel}},
  \bibinfo {author} {\bibfnamefont {A.}~\bibnamefont {Rycerz}}, \bibinfo
  {author} {\bibfnamefont {Y.~M.}\ \bibnamefont {Blanter}}, \bibinfo {author}
  {\bibfnamefont {C.~W.~J.}\ \bibnamefont {Beenakker}}, \ and\ \bibinfo
  {author} {\bibfnamefont {A.~F.}\ \bibnamefont {Morpurgo}},\ }\href@noop {}
  {\bibfield  {journal} {\bibinfo  {journal} {Phys. Rev. B}\ }\textbf {\bibinfo
  {volume} {76}},\ \bibinfo {pages} {235404} (\bibinfo {year}
  {2007})}\BibitemShut {NoStop}%
\bibitem [{\citenamefont {Jackiw}\ \emph {et~al.}(2009)\citenamefont {Jackiw},
  \citenamefont {Milstein}, \citenamefont {Pi},\ and\ \citenamefont
  {Terekhov}}]{tg2}%
  \BibitemOpen
  \bibfield  {author} {\bibinfo {author} {\bibfnamefont {R.}~\bibnamefont
  {Jackiw}}, \bibinfo {author} {\bibfnamefont {A.~I.}\ \bibnamefont
  {Milstein}}, \bibinfo {author} {\bibfnamefont {S.-Y.}\ \bibnamefont {Pi}}, \
  and\ \bibinfo {author} {\bibfnamefont {I.~S.}\ \bibnamefont {Terekhov}},\
  }\href@noop {} {\bibfield  {journal} {\bibinfo  {journal} {Phys. Rev. B}\
  }\textbf {\bibinfo {volume} {80}},\ \bibinfo {pages} {033413} (\bibinfo
  {year} {2009})}\BibitemShut {NoStop}%
\bibitem [{\citenamefont {Schelter}\ \emph {et~al.}(2010)\citenamefont
  {Schelter}, \citenamefont {Bohr},\ and\ \citenamefont {Trauzettel}}]{tg3}%
  \BibitemOpen
  \bibfield  {author} {\bibinfo {author} {\bibfnamefont {J.}~\bibnamefont
  {Schelter}}, \bibinfo {author} {\bibfnamefont {D.}~\bibnamefont {Bohr}}, \
  and\ \bibinfo {author} {\bibfnamefont {B.}~\bibnamefont {Trauzettel}},\
  }\href@noop {} {\bibfield  {journal} {\bibinfo  {journal} {Phys. Rev. B}\
  }\textbf {\bibinfo {volume} {81}},\ \bibinfo {pages} {195441} (\bibinfo
  {year} {2010})}\BibitemShut {NoStop}%
\bibitem [{\citenamefont {Faria}\ \emph {et~al.}(2013)\citenamefont {Faria},
  \citenamefont {Latg\'e}, \citenamefont {Ulloa},\ and\ \citenamefont
  {Sandler}}]{tg4}%
  \BibitemOpen
  \bibfield  {author} {\bibinfo {author} {\bibfnamefont {D.}~\bibnamefont
  {Faria}}, \bibinfo {author} {\bibfnamefont {A.}~\bibnamefont {Latg\'e}},
  \bibinfo {author} {\bibfnamefont {S.~E.}\ \bibnamefont {Ulloa}}, \ and\
  \bibinfo {author} {\bibfnamefont {N.}~\bibnamefont {Sandler}},\ }\href@noop
  {} {\bibfield  {journal} {\bibinfo  {journal} {Phys. Rev. B}\ }\textbf
  {\bibinfo {volume} {87}},\ \bibinfo {pages} {241403} (\bibinfo {year}
  {2013})}\BibitemShut {NoStop}%
\bibitem [{\citenamefont {Wurm}\ \emph {et~al.}(2010)\citenamefont {Wurm},
  \citenamefont {Wimmer}, \citenamefont {Baranger},\ and\ \citenamefont
  {Richter}}]{tg5}%
  \BibitemOpen
  \bibfield  {author} {\bibinfo {author} {\bibfnamefont {J.}~\bibnamefont
  {Wurm}}, \bibinfo {author} {\bibfnamefont {M.}~\bibnamefont {Wimmer}},
  \bibinfo {author} {\bibfnamefont {H.~U.}\ \bibnamefont {Baranger}}, \ and\
  \bibinfo {author} {\bibfnamefont {K.}~\bibnamefont {Richter}},\ }\href@noop
  {} {\bibfield  {journal} {\bibinfo  {journal} {Semicond. Sci. Technol.}\
  }\textbf {\bibinfo {volume} {25}},\ \bibinfo {pages} {034003} (\bibinfo
  {year} {2010})}\BibitemShut {NoStop}%
\bibitem [{\citenamefont {Wu}\ \emph {et~al.}(2010)\citenamefont {Wu},
  \citenamefont {Zhang}, \citenamefont {Chang},\ and\ \citenamefont
  {Peeters}}]{tg6}%
  \BibitemOpen
  \bibfield  {author} {\bibinfo {author} {\bibfnamefont {Z.}~\bibnamefont
  {Wu}}, \bibinfo {author} {\bibfnamefont {Z.~Z.}\ \bibnamefont {Zhang}},
  \bibinfo {author} {\bibfnamefont {K.}~\bibnamefont {Chang}}, \ and\ \bibinfo
  {author} {\bibfnamefont {F.~M.}\ \bibnamefont {Peeters}},\ }\href
  {http://stacks.iop.org/0957-4484/21/i=18/a=185201} {\bibfield  {journal}
  {\bibinfo  {journal} {Nanotechnology}\ }\textbf {\bibinfo {volume} {21}},\
  \bibinfo {pages} {185201} (\bibinfo {year} {2010})}\BibitemShut {NoStop}%
\bibitem [{\citenamefont {Abergel}\ \emph {et~al.}(2008)\citenamefont
  {Abergel}, \citenamefont {Apalkov},\ and\ \citenamefont {Chakraborty}}]{tg8}%
  \BibitemOpen
  \bibfield  {author} {\bibinfo {author} {\bibfnamefont {D.~S.~L.}\
  \bibnamefont {Abergel}}, \bibinfo {author} {\bibfnamefont {V.~M.}\
  \bibnamefont {Apalkov}}, \ and\ \bibinfo {author} {\bibfnamefont
  {T.}~\bibnamefont {Chakraborty}},\ }\href@noop {} {\bibfield  {journal}
  {\bibinfo  {journal} {Phys. Rev. B}\ }\textbf {\bibinfo {volume} {78}},\
  \bibinfo {pages} {193405} (\bibinfo {year} {2008})}\BibitemShut {NoStop}%
\bibitem [{\citenamefont {da~Costa}\ \emph {et~al.}(2014)\citenamefont
  {da~Costa}, \citenamefont {Chaves}, \citenamefont {Zarenia}, \citenamefont
  {Pereira}, \citenamefont {Farias},\ and\ \citenamefont {Peeters}}]{tg9}%
  \BibitemOpen
  \bibfield  {author} {\bibinfo {author} {\bibfnamefont {D.~R.}\ \bibnamefont
  {da~Costa}}, \bibinfo {author} {\bibfnamefont {A.}~\bibnamefont {Chaves}},
  \bibinfo {author} {\bibfnamefont {M.}~\bibnamefont {Zarenia}}, \bibinfo
  {author} {\bibfnamefont {J.~M.}\ \bibnamefont {Pereira}}, \bibinfo {author}
  {\bibfnamefont {G.~A.}\ \bibnamefont {Farias}}, \ and\ \bibinfo {author}
  {\bibfnamefont {F.~M.}\ \bibnamefont {Peeters}},\ }\href@noop {} {\bibfield
  {journal} {\bibinfo  {journal} {Phys. Rev. B}\ }\textbf {\bibinfo {volume}
  {89}},\ \bibinfo {pages} {075418} (\bibinfo {year} {2014})}\BibitemShut
  {NoStop}%
\bibitem [{\citenamefont {Hewageegana}\ and\ \citenamefont
  {Apalkov}(2008)}]{tg10}%
  \BibitemOpen
  \bibfield  {author} {\bibinfo {author} {\bibfnamefont {P.}~\bibnamefont
  {Hewageegana}}\ and\ \bibinfo {author} {\bibfnamefont {V.}~\bibnamefont
  {Apalkov}},\ }\href@noop {} {\bibfield  {journal} {\bibinfo  {journal} {Phys.
  Rev. B}\ }\textbf {\bibinfo {volume} {77}},\ \bibinfo {pages} {245426}
  (\bibinfo {year} {2008})}\BibitemShut {NoStop}%
\bibitem [{\citenamefont {Downing}\ \emph {et~al.}(2011)\citenamefont
  {Downing}, \citenamefont {Stone},\ and\ \citenamefont {Portnoi}}]{tg11}%
  \BibitemOpen
  \bibfield  {author} {\bibinfo {author} {\bibfnamefont {C.~A.}\ \bibnamefont
  {Downing}}, \bibinfo {author} {\bibfnamefont {D.~A.}\ \bibnamefont {Stone}},
  \ and\ \bibinfo {author} {\bibfnamefont {M.~E.}\ \bibnamefont {Portnoi}},\
  }\href {\doibase 10.1103/PhysRevB.84.155437} {\bibfield  {journal} {\bibinfo
  {journal} {Phys. Rev. B}\ }\textbf {\bibinfo {volume} {84}},\ \bibinfo
  {pages} {155437} (\bibinfo {year} {2011})}\BibitemShut {NoStop}%
\bibitem [{\citenamefont {Rakyta}\ \emph {et~al.}(2014)\citenamefont {Rakyta},
  \citenamefont {T\'ov\'ari}, \citenamefont {Csontos}, \citenamefont {Csonka},
  \citenamefont {Csord\'as},\ and\ \citenamefont {Cserti}}]{tg13}%
  \BibitemOpen
  \bibfield  {author} {\bibinfo {author} {\bibfnamefont {P.}~\bibnamefont
  {Rakyta}}, \bibinfo {author} {\bibfnamefont {E.}~\bibnamefont {T\'ov\'ari}},
  \bibinfo {author} {\bibfnamefont {M.}~\bibnamefont {Csontos}}, \bibinfo
  {author} {\bibfnamefont {S.}~\bibnamefont {Csonka}}, \bibinfo {author}
  {\bibfnamefont {A.}~\bibnamefont {Csord\'as}}, \ and\ \bibinfo {author}
  {\bibfnamefont {J.}~\bibnamefont {Cserti}},\ }\href@noop {} {\bibfield
  {journal} {\bibinfo  {journal} {Phys. Rev. B}\ }\textbf {\bibinfo {volume}
  {90}},\ \bibinfo {pages} {125428} (\bibinfo {year} {2014})}\BibitemShut
  {NoStop}%
\bibitem [{\citenamefont {Mandelshtam}\ \emph {et~al.}(1993)\citenamefont
  {Mandelshtam}, \citenamefont {Ravuri},\ and\ \citenamefont
  {Taylor}}]{mandelsh}%
  \BibitemOpen
  \bibfield  {author} {\bibinfo {author} {\bibfnamefont {V.~A.}\ \bibnamefont
  {Mandelshtam}}, \bibinfo {author} {\bibfnamefont {T.~R.}\ \bibnamefont
  {Ravuri}}, \ and\ \bibinfo {author} {\bibfnamefont {H.~S.}\ \bibnamefont
  {Taylor}},\ }\href {\doibase 10.1103/PhysRevLett.70.1932} {\bibfield
  {journal} {\bibinfo  {journal} {Phys. Rev. Lett.}\ }\textbf {\bibinfo
  {volume} {70}},\ \bibinfo {pages} {1932} (\bibinfo {year}
  {1993})}\BibitemShut {NoStop}%
\bibitem [{\citenamefont {Hansen}\ \emph {et~al.}(2001)\citenamefont {Hansen},
  \citenamefont {Kristensen}, \citenamefont {Pedersen}, \citenamefont
  {S\o{}rensen},\ and\ \citenamefont {Lindelof}}]{12}%
  \BibitemOpen
  \bibfield  {author} {\bibinfo {author} {\bibfnamefont {A.~E.}\ \bibnamefont
  {Hansen}}, \bibinfo {author} {\bibfnamefont {A.}~\bibnamefont {Kristensen}},
  \bibinfo {author} {\bibfnamefont {S.}~\bibnamefont {Pedersen}}, \bibinfo
  {author} {\bibfnamefont {C.~B.}\ \bibnamefont {S\o{}rensen}}, \ and\ \bibinfo
  {author} {\bibfnamefont {P.~E.}\ \bibnamefont {Lindelof}},\ }\href {\doibase
  10.1103/PhysRevB.64.045327} {\bibfield  {journal} {\bibinfo  {journal} {Phys.
  Rev. B}\ }\textbf {\bibinfo {volume} {64}},\ \bibinfo {pages} {045327}
  (\bibinfo {year} {2001})}\BibitemShut {NoStop}%
\bibitem [{\citenamefont {Grbi\'c}\ \emph {et~al.}(2008)\citenamefont
  {Grbi\'c}, \citenamefont {Leturcq}, \citenamefont {Ihn}, \citenamefont
  {Ensslin}, \citenamefont {Reuter},\ and\ \citenamefont {Wieck}}]{13}%
  \BibitemOpen
  \bibfield  {author} {\bibinfo {author} {\bibfnamefont {B.}~\bibnamefont
  {Grbi\'c}}, \bibinfo {author} {\bibfnamefont {R.}~\bibnamefont {Leturcq}},
  \bibinfo {author} {\bibfnamefont {T.}~\bibnamefont {Ihn}}, \bibinfo {author}
  {\bibfnamefont {K.}~\bibnamefont {Ensslin}}, \bibinfo {author} {\bibfnamefont
  {D.}~\bibnamefont {Reuter}}, \ and\ \bibinfo {author} {\bibfnamefont {A.~D.}\
  \bibnamefont {Wieck}},\ }\href {\doibase
  http://dx.doi.org/10.1016/j.physe.2007.08.129} {\bibfield  {journal}
  {\bibinfo  {journal} {Physica E: Low-dimensional Systems and Nanostructures}\
  }\textbf {\bibinfo {volume} {40}},\ \bibinfo {pages} {1273 } (\bibinfo {year}
  {2008})},\ \bibinfo {note} {17th International Conference on Electronic
  Properties of Two-Dimensional Systems}\BibitemShut {NoStop}%
\bibitem [{\citenamefont {Al'tshuler}\ \emph {et~al.}(1981)\citenamefont
  {Al'tshuler}, \citenamefont {Aronov},\ and\ \citenamefont {Spivak}}]{spivak}%
  \BibitemOpen
  \bibfield  {author} {\bibinfo {author} {\bibfnamefont {B.~L.}\ \bibnamefont
  {Al'tshuler}}, \bibinfo {author} {\bibfnamefont {A.~G.}\ \bibnamefont
  {Aronov}}, \ and\ \bibinfo {author} {\bibfnamefont {B.~Z.}\ \bibnamefont
  {Spivak}},\ }\href@noop {} {\bibfield  {journal} {\bibinfo  {journal} {JETP
  Lett.}\ }\textbf {\bibinfo {volume} {33}},\ \bibinfo {pages} {94} (\bibinfo
  {year} {1981})}\BibitemShut {NoStop}%
\bibitem [{\citenamefont {D'Anjou}\ and\ \citenamefont {Coish}(2013)}]{Anjou}%
  \BibitemOpen
  \bibfield  {author} {\bibinfo {author} {\bibfnamefont {B.}~\bibnamefont
  {D'Anjou}}\ and\ \bibinfo {author} {\bibfnamefont {W.~A.}\ \bibnamefont
  {Coish}},\ }\href {\doibase 10.1103/PhysRevB.87.085443} {\bibfield  {journal}
  {\bibinfo  {journal} {Phys. Rev. B}\ }\textbf {\bibinfo {volume} {87}},\
  \bibinfo {pages} {085443} (\bibinfo {year} {2013})}\BibitemShut {NoStop}%
\bibitem [{\citenamefont {Kubo}\ \emph {et~al.}(2011)\citenamefont {Kubo},
  \citenamefont {Ichigo},\ and\ \citenamefont {Tokura}}]{Kubo}%
  \BibitemOpen
  \bibfield  {author} {\bibinfo {author} {\bibfnamefont {T.}~\bibnamefont
  {Kubo}}, \bibinfo {author} {\bibfnamefont {Y.}~\bibnamefont {Ichigo}}, \ and\
  \bibinfo {author} {\bibfnamefont {Y.}~\bibnamefont {Tokura}},\ }\href
  {\doibase 10.1103/PhysRevB.83.235310} {\bibfield  {journal} {\bibinfo
  {journal} {Phys. Rev. B}\ }\textbf {\bibinfo {volume} {83}},\ \bibinfo
  {pages} {235310} (\bibinfo {year} {2011})}\BibitemShut {NoStop}%
\bibitem [{\citenamefont {Bedkihal}\ \emph {et~al.}(2013)\citenamefont
  {Bedkihal}, \citenamefont {Bandyopadhyay},\ and\ \citenamefont
  {Segal}}]{Bedkihal}%
  \BibitemOpen
  \bibfield  {author} {\bibinfo {author} {\bibfnamefont {S.}~\bibnamefont
  {Bedkihal}}, \bibinfo {author} {\bibfnamefont {M.}~\bibnamefont
  {Bandyopadhyay}}, \ and\ \bibinfo {author} {\bibfnamefont {D.}~\bibnamefont
  {Segal}},\ }\href {\doibase 10.1103/PhysRevB.87.045418} {\bibfield  {journal}
  {\bibinfo  {journal} {Phys. Rev. B}\ }\textbf {\bibinfo {volume} {87}},\
  \bibinfo {pages} {045418} (\bibinfo {year} {2013})}\BibitemShut {NoStop}%
\bibitem [{\citenamefont {Carmier}\ \emph {et~al.}(2010)\citenamefont
  {Carmier}, \citenamefont {Lewenkopf},\ and\ \citenamefont {Ullmo}}]{snakes}%
  \BibitemOpen
  \bibfield  {author} {\bibinfo {author} {\bibfnamefont {P.}~\bibnamefont
  {Carmier}}, \bibinfo {author} {\bibfnamefont {C.}~\bibnamefont {Lewenkopf}},
  \ and\ \bibinfo {author} {\bibfnamefont {D.}~\bibnamefont {Ullmo}},\ }\href
  {\doibase 10.1103/PhysRevB.81.241406} {\bibfield  {journal} {\bibinfo
  {journal} {Phys. Rev. B}\ }\textbf {\bibinfo {volume} {81}},\ \bibinfo
  {pages} {241406(R)} (\bibinfo {year} {2010})}\BibitemShut {NoStop}%
\bibitem [{\citenamefont {Patel}\ \emph {et~al.}(2012)\citenamefont {Patel},
  \citenamefont {Davies}, \citenamefont {Cheianov},\ and\ \citenamefont
  {Fal'ko}}]{snakes2}%
  \BibitemOpen
  \bibfield  {author} {\bibinfo {author} {\bibfnamefont {A.~A.}\ \bibnamefont
  {Patel}}, \bibinfo {author} {\bibfnamefont {N.}~\bibnamefont {Davies}},
  \bibinfo {author} {\bibfnamefont {V.}~\bibnamefont {Cheianov}}, \ and\
  \bibinfo {author} {\bibfnamefont {V.~I.}\ \bibnamefont {Fal'ko}},\ }\href
  {\doibase 10.1103/PhysRevB.86.081413} {\bibfield  {journal} {\bibinfo
  {journal} {Phys. Rev. B}\ }\textbf {\bibinfo {volume} {86}},\ \bibinfo
  {pages} {081413} (\bibinfo {year} {2012})}\BibitemShut {NoStop}%
\bibitem [{\citenamefont {Zarenia}\ \emph {et~al.}(2013)\citenamefont
  {Zarenia}, \citenamefont {Pereira}, \citenamefont {Peeters},\ and\
  \citenamefont {Farias}}]{snakes3}%
  \BibitemOpen
  \bibfield  {author} {\bibinfo {author} {\bibfnamefont {M.}~\bibnamefont
  {Zarenia}}, \bibinfo {author} {\bibfnamefont {J.~M.}\ \bibnamefont
  {Pereira}}, \bibinfo {author} {\bibfnamefont {F.~M.}\ \bibnamefont
  {Peeters}}, \ and\ \bibinfo {author} {\bibfnamefont {G.~A.}\ \bibnamefont
  {Farias}},\ }\href@noop {} {\bibfield  {journal} {\bibinfo  {journal} {Phys.
  Rev. B}\ }\textbf {\bibinfo {volume} {87}},\ \bibinfo {pages} {035426}
  (\bibinfo {year} {2013})}\BibitemShut {NoStop}%
\bibitem [{\citenamefont {Milovanovi\'{c}}\ \emph {et~al.}(2014)\citenamefont
  {Milovanovi\'{c}}, \citenamefont {Ramezani~Masir},\ and\ \citenamefont
  {Peeters}}]{snakes4}%
  \BibitemOpen
  \bibfield  {author} {\bibinfo {author} {\bibfnamefont {S.~P.}\ \bibnamefont
  {Milovanovi\'{c}}}, \bibinfo {author} {\bibfnamefont {M.}~\bibnamefont
  {Ramezani~Masir}}, \ and\ \bibinfo {author} {\bibfnamefont {F.~M.}\
  \bibnamefont {Peeters}},\ }\href
  {http://scitation.aip.org/content/aip/journal/apl/105/12/10.1063/1.4896769}
  {\bibfield  {journal} {\bibinfo  {journal} {Appl. Phys. Lett.}\ }\textbf
  {\bibinfo {volume} {105}},\ \bibinfo {eid} {123507} (\bibinfo {year}
  {2014})}\BibitemShut {NoStop}%
\bibitem [{\citenamefont {{Cohnitz}}\ \emph {et~al.}(2016)\citenamefont
  {{Cohnitz}}, \citenamefont {{De Martino}}, \citenamefont {{H{\"a}usler}},\
  and\ \citenamefont {{Egger}}}]{snakes5}%
  \BibitemOpen
  \bibfield  {author} {\bibinfo {author} {\bibfnamefont {L.}~\bibnamefont
  {{Cohnitz}}}, \bibinfo {author} {\bibfnamefont {A.}~\bibnamefont {{De
  Martino}}}, \bibinfo {author} {\bibfnamefont {W.}~\bibnamefont
  {{H{\"a}usler}}}, \ and\ \bibinfo {author} {\bibfnamefont {R.}~\bibnamefont
  {{Egger}}},\ }\href@noop {} {\  (\bibinfo {year} {2016})},\ \Eprint
  {http://arxiv.org/abs/1608.03469} {arXiv:1608.03469 [cond-mat.mes-hall]}
  \BibitemShut {NoStop}%
\bibitem [{\citenamefont {Rickhaus}\ \emph {et~al.}(2015)\citenamefont
  {Rickhaus}, \citenamefont {Makk}, \citenamefont {Liu}, \citenamefont
  {T\'{o}v\'{a}ri}, \citenamefont {Weiss}, \citenamefont {Maurand},
  \citenamefont {Richter},\ and\ \citenamefont {Sch\"onenberger}}]{snakes6}%
  \BibitemOpen
  \bibfield  {author} {\bibinfo {author} {\bibfnamefont {P.}~\bibnamefont
  {Rickhaus}}, \bibinfo {author} {\bibfnamefont {P.}~\bibnamefont {Makk}},
  \bibinfo {author} {\bibfnamefont {M.-H.}\ \bibnamefont {Liu}}, \bibinfo
  {author} {\bibfnamefont {E.}~\bibnamefont {T\'{o}v\'{a}ri}}, \bibinfo
  {author} {\bibfnamefont {M.}~\bibnamefont {Weiss}}, \bibinfo {author}
  {\bibfnamefont {R.}~\bibnamefont {Maurand}}, \bibinfo {author} {\bibfnamefont
  {K.}~\bibnamefont {Richter}}, \ and\ \bibinfo {author} {\bibfnamefont
  {C.}~\bibnamefont {Sch\"onenberger}},\ }\href@noop {} {\bibfield  {journal}
  {\bibinfo  {journal} {Nat. Commun.}\ }\textbf {\bibinfo {volume} {6}},\
  \bibinfo {eid} {6470} (\bibinfo {year} {2015})}\BibitemShut {NoStop}%
\bibitem [{\citenamefont {Ghosh}\ \emph {et~al.}(2008)\citenamefont {Ghosh},
  \citenamefont {De~Martino}, \citenamefont {H\"ausler}, \citenamefont
  {Dell'Anna},\ and\ \citenamefont {Egger}}]{Ghosh}%
  \BibitemOpen
  \bibfield  {author} {\bibinfo {author} {\bibfnamefont {T.~K.}\ \bibnamefont
  {Ghosh}}, \bibinfo {author} {\bibfnamefont {A.}~\bibnamefont {De~Martino}},
  \bibinfo {author} {\bibfnamefont {W.}~\bibnamefont {H\"ausler}}, \bibinfo
  {author} {\bibfnamefont {L.}~\bibnamefont {Dell'Anna}}, \ and\ \bibinfo
  {author} {\bibfnamefont {R.}~\bibnamefont {Egger}},\ }\href {\doibase
  10.1103/PhysRevB.77.081404} {\bibfield  {journal} {\bibinfo  {journal} {Phys.
  Rev. B}\ }\textbf {\bibinfo {volume} {77}},\ \bibinfo {pages} {081404(R)}
  (\bibinfo {year} {2008})}\BibitemShut {NoStop}%
\bibitem [{\citenamefont {Cresti}\ \emph {et~al.}(2008)\citenamefont {Cresti},
  \citenamefont {Grosso},\ and\ \citenamefont {Parravicini}}]{Cresti}%
  \BibitemOpen
  \bibfield  {author} {\bibinfo {author} {\bibfnamefont {A.}~\bibnamefont
  {Cresti}}, \bibinfo {author} {\bibfnamefont {G.}~\bibnamefont {Grosso}}, \
  and\ \bibinfo {author} {\bibfnamefont {G.~P.}\ \bibnamefont {Parravicini}},\
  }\href@noop {} {\bibfield  {journal} {\bibinfo  {journal} {Phys. Rev. B}\
  }\textbf {\bibinfo {volume} {77}},\ \bibinfo {pages} {115408} (\bibinfo
  {year} {2008})}\BibitemShut {NoStop}%
\bibitem [{\citenamefont {Rickhaus}\ \emph {et~al.}(2013)\citenamefont
  {Rickhaus}, \citenamefont {Maurand}, \citenamefont {Liu}, \citenamefont
  {Weiss}, \citenamefont {Richter},\ and\ \citenamefont
  {Sch{\"o}nenberger}}]{Rickhaus}%
  \BibitemOpen
  \bibfield  {author} {\bibinfo {author} {\bibfnamefont {P.}~\bibnamefont
  {Rickhaus}}, \bibinfo {author} {\bibfnamefont {R.}~\bibnamefont {Maurand}},
  \bibinfo {author} {\bibfnamefont {M.-H.}\ \bibnamefont {Liu}}, \bibinfo
  {author} {\bibfnamefont {M.}~\bibnamefont {Weiss}}, \bibinfo {author}
  {\bibfnamefont {K.}~\bibnamefont {Richter}}, \ and\ \bibinfo {author}
  {\bibfnamefont {C.}~\bibnamefont {Sch{\"o}nenberger}},\ }\href@noop {}
  {\bibfield  {journal} {\bibinfo  {journal} {Nat. Commun.}\ }\textbf {\bibinfo
  {volume} {4}},\ \bibinfo {eid} {2342} (\bibinfo {year} {2013})}\BibitemShut
  {NoStop}%
\bibitem [{\citenamefont {Liu}\ \emph {et~al.}(2015{\natexlab{a}})\citenamefont
  {Liu}, \citenamefont {Tiwari}, \citenamefont {Brada}, \citenamefont {Bruder},
  \citenamefont {Kusmartsev},\ and\ \citenamefont {Mele}}]{Yliu}%
  \BibitemOpen
  \bibfield  {author} {\bibinfo {author} {\bibfnamefont {Y.}~\bibnamefont
  {Liu}}, \bibinfo {author} {\bibfnamefont {R.~P.}\ \bibnamefont {Tiwari}},
  \bibinfo {author} {\bibfnamefont {M.}~\bibnamefont {Brada}}, \bibinfo
  {author} {\bibfnamefont {C.}~\bibnamefont {Bruder}}, \bibinfo {author}
  {\bibfnamefont {F.~V.}\ \bibnamefont {Kusmartsev}}, \ and\ \bibinfo {author}
  {\bibfnamefont {E.~J.}\ \bibnamefont {Mele}},\ }\href {\doibase
  10.1103/PhysRevB.92.235438} {\bibfield  {journal} {\bibinfo  {journal} {Phys.
  Rev. B}\ }\textbf {\bibinfo {volume} {92}},\ \bibinfo {pages} {235438}
  (\bibinfo {year} {2015}{\natexlab{a}})}\BibitemShut {NoStop}%
\bibitem [{\citenamefont {Mre\'{n}ca-Kolasi\'{n}ska}\ \emph
  {et~al.}(2016)\citenamefont {Mre\'{n}ca-Kolasi\'{n}ska}, \citenamefont
  {Heun},\ and\ \citenamefont {Szafran}}]{amrenca}%
  \BibitemOpen
  \bibfield  {author} {\bibinfo {author} {\bibfnamefont {A.}~\bibnamefont
  {Mre\'{n}ca-Kolasi\'{n}ska}}, \bibinfo {author} {\bibfnamefont
  {S.}~\bibnamefont {Heun}}, \ and\ \bibinfo {author} {\bibfnamefont
  {B.}~\bibnamefont {Szafran}},\ }\href {\doibase 10.1103/PhysRevB.93.125411}
  {\bibfield  {journal} {\bibinfo  {journal} {Phys. Rev. B}\ }\textbf {\bibinfo
  {volume} {93}},\ \bibinfo {pages} {125411} (\bibinfo {year}
  {2016})}\BibitemShut {NoStop}%
\bibitem [{\citenamefont {Kolasi\'{n}ski}\ and\ \citenamefont
  {Szafran}(2013)}]{kolasinskiDFT2013}%
  \BibitemOpen
  \bibfield  {author} {\bibinfo {author} {\bibfnamefont {K.}~\bibnamefont
  {Kolasi\'{n}ski}}\ and\ \bibinfo {author} {\bibfnamefont {B.}~\bibnamefont
  {Szafran}},\ }\href {\doibase 10.1103/PhysRevB.88.165306} {\bibfield
  {journal} {\bibinfo  {journal} {Phys. Rev. B}\ }\textbf {\bibinfo {volume}
  {88}},\ \bibinfo {pages} {165306} (\bibinfo {year} {2013})}\BibitemShut
  {NoStop}%
\bibitem [{\citenamefont {Liu}\ \emph {et~al.}(2015{\natexlab{b}})\citenamefont
  {Liu}, \citenamefont {Rickhaus}, \citenamefont {Makk}, \citenamefont
  {T\'ov\'ari}, \citenamefont {Maurand}, \citenamefont {Tkatschenko},
  \citenamefont {Weiss}, \citenamefont {Sch\"onenberger},\ and\ \citenamefont
  {Richter}}]{Rickhausprl}%
  \BibitemOpen
  \bibfield  {author} {\bibinfo {author} {\bibfnamefont {M.-H.}\ \bibnamefont
  {Liu}}, \bibinfo {author} {\bibfnamefont {P.}~\bibnamefont {Rickhaus}},
  \bibinfo {author} {\bibfnamefont {P.}~\bibnamefont {Makk}}, \bibinfo {author}
  {\bibfnamefont {E.}~\bibnamefont {T\'ov\'ari}}, \bibinfo {author}
  {\bibfnamefont {R.}~\bibnamefont {Maurand}}, \bibinfo {author} {\bibfnamefont
  {F.}~\bibnamefont {Tkatschenko}}, \bibinfo {author} {\bibfnamefont
  {M.}~\bibnamefont {Weiss}}, \bibinfo {author} {\bibfnamefont
  {C.}~\bibnamefont {Sch\"onenberger}}, \ and\ \bibinfo {author} {\bibfnamefont
  {K.}~\bibnamefont {Richter}},\ }\href {\doibase
  10.1103/PhysRevLett.114.036601} {\bibfield  {journal} {\bibinfo  {journal}
  {Phys. Rev. Lett.}\ }\textbf {\bibinfo {volume} {114}},\ \bibinfo {pages}
  {036601} (\bibinfo {year} {2015}{\natexlab{b}})}\BibitemShut {NoStop}%
\bibitem [{\citenamefont {Kolasi\'{n}ski}\ \emph {et~al.}(2016)\citenamefont
  {Kolasi\'{n}ski}, \citenamefont {Szafran}, \citenamefont {Brun},\ and\
  \citenamefont {Sellier}}]{Kolacha}%
  \BibitemOpen
  \bibfield  {author} {\bibinfo {author} {\bibfnamefont {K.}~\bibnamefont
  {Kolasi\'{n}ski}}, \bibinfo {author} {\bibfnamefont {B.}~\bibnamefont
  {Szafran}}, \bibinfo {author} {\bibfnamefont {B.}~\bibnamefont {Brun}}, \
  and\ \bibinfo {author} {\bibfnamefont {H.}~\bibnamefont {Sellier}},\ }\href
  {\doibase 10.1103/PhysRevB.94.075301} {\bibfield  {journal} {\bibinfo
  {journal} {Phys. Rev. B}\ }\textbf {\bibinfo {volume} {94}},\ \bibinfo
  {pages} {075301} (\bibinfo {year} {2016})}\BibitemShut {NoStop}%
\bibitem [{\citenamefont {Datta}(1997)}]{datta}%
  \BibitemOpen
  \bibfield  {author} {\bibinfo {author} {\bibfnamefont {S.}~\bibnamefont
  {Datta}},\ }\href@noop {} {\emph {\bibinfo {title} {Electronic Transport in
  Mesoscopic Systems}}}\ (\bibinfo  {publisher} {Cambridge University Press},\
  \bibinfo {year} {1997})\BibitemShut {NoStop}%
\bibitem [{\citenamefont {Wakabayashi}(2001)}]{Wakabayashi}%
  \BibitemOpen
  \bibfield  {author} {\bibinfo {author} {\bibfnamefont {K.}~\bibnamefont
  {Wakabayashi}},\ }\href {\doibase 10.1103/PhysRevB.64.125428} {\bibfield
  {journal} {\bibinfo  {journal} {Phys. Rev. B}\ }\textbf {\bibinfo {volume}
  {64}},\ \bibinfo {pages} {125428} (\bibinfo {year} {2001})}\BibitemShut
  {NoStop}%
\bibitem [{uwa()}]{uwaga}%
  \BibitemOpen
  \href@noop {} {\ }\bibinfo {note} {The unequal currents in the upper and
  lower arms of the ring in Fig.~\ref{strum}(a) for $B=0$ result from an
  asymmetry of the device which occurs at the atomic level near the ring-ribbon
  junction.}\BibitemShut {Stop}%
\bibitem [{\citenamefont {Szafran}(2008)}]{szafran}%
  \BibitemOpen
  \bibfield  {author} {\bibinfo {author} {\bibfnamefont {B.}~\bibnamefont
  {Szafran}},\ }\href {\doibase 10.1103/PhysRevB.77.205313} {\bibfield
  {journal} {\bibinfo  {journal} {Phys. Rev. B}\ }\textbf {\bibinfo {volume}
  {77}},\ \bibinfo {pages} {205313} (\bibinfo {year} {2008})}\BibitemShut
  {NoStop}%
\bibitem [{\citenamefont {Wakabayashi}\ \emph {et~al.}(2007)\citenamefont
  {Wakabayashi}, \citenamefont {Takane},\ and\ \citenamefont {Sigrist}}]{pcc}%
  \BibitemOpen
  \bibfield  {author} {\bibinfo {author} {\bibfnamefont {K.}~\bibnamefont
  {Wakabayashi}}, \bibinfo {author} {\bibfnamefont {Y.}~\bibnamefont {Takane}},
  \ and\ \bibinfo {author} {\bibfnamefont {M.}~\bibnamefont {Sigrist}},\ }\href
  {\doibase 10.1103/PhysRevLett.99.036601} {\bibfield  {journal} {\bibinfo
  {journal} {Phys. Rev. Lett.}\ }\textbf {\bibinfo {volume} {99}},\ \bibinfo
  {pages} {036601} (\bibinfo {year} {2007})}\BibitemShut {NoStop}%
\bibitem [{\citenamefont {Wakabayashi}\ \emph {et~al.}(2009)\citenamefont
  {Wakabayashi}, \citenamefont {Takane}, \citenamefont {Yamamoto},\ and\
  \citenamefont {Sigrist}}]{pccc}%
  \BibitemOpen
  \bibfield  {author} {\bibinfo {author} {\bibfnamefont {K.}~\bibnamefont
  {Wakabayashi}}, \bibinfo {author} {\bibfnamefont {Y.}~\bibnamefont {Takane}},
  \bibinfo {author} {\bibfnamefont {M.}~\bibnamefont {Yamamoto}}, \ and\
  \bibinfo {author} {\bibfnamefont {M.}~\bibnamefont {Sigrist}},\ }\href
  {\doibase http://dx.doi.org/10.1016/j.carbon.2008.09.040} {\bibfield
  {journal} {\bibinfo  {journal} {Carbon}\ }\textbf {\bibinfo {volume} {47}},\
  \bibinfo {pages} {124 } (\bibinfo {year} {2009})}\BibitemShut {NoStop}%
\bibitem [{\citenamefont {Poniedzia\l{}ek}\ and\ \citenamefont
  {Szafran}(2010)}]{Poniedzialek}%
  \BibitemOpen
  \bibfield  {author} {\bibinfo {author} {\bibfnamefont {M.~R.}\ \bibnamefont
  {Poniedzia\l{}ek}}\ and\ \bibinfo {author} {\bibfnamefont {B.}~\bibnamefont
  {Szafran}},\ }\href {http://stacks.iop.org/0953-8984/22/i=46/a=465801}
  {\bibfield  {journal} {\bibinfo  {journal} {J. Phys.: Condens. Matter}\
  }\textbf {\bibinfo {volume} {22}},\ \bibinfo {pages} {465801} (\bibinfo
  {year} {2010})}\BibitemShut {NoStop}%
\end{thebibliography}%

\end{document}